\documentclass[preprint,amsart,showpacs,preprintnumbers,amsmath,amssymb]{revtex4-1}
\usepackage[T1]{fontenc}
\usepackage[latin9]{inputenc}
\usepackage[english]{babel}
\usepackage{graphicx}
\usepackage{dcolumn}
\usepackage{bm}
\usepackage{bbm}
\usepackage{hyperref}
\usepackage{color}
\usepackage{mathtools}

\begin{document}
\title{The nature of domain walls in ultrathin ferromagnets\\
 revealed by scanning nanomagnetometry}
\author{J.-P.~Tetienne$^{1,\dag}$, T. Hingant$^{1,\dag}$, L. J. Martinez$^{1}$, S.~Rohart$^2$, A.~Thiaville$^2$, L.~Herrera~Diez$^{3}$, K.~Garcia$^{3}$, J.-P.~Adam$^{3}$, J.-V.~Kim$^{3}$, J.-F. Roch$^{1}$, I. M. Miron$^{4}$, G. Gaudin$^{4}$, L. Vila$^{5}$, B. Ocker$^{6}$, D.~Ravelosona$^3$ and V.~Jacques$^{1}$}
\email{vjacques@ens-cachan.fr}
\affiliation{$^{1}$Laboratoire Aim\'e Cotton, CNRS, Universit\'e Paris-Sud and ENS Cachan, 91405 Orsay, France \\
$^{2}$Laboratoire de Physique des Solides, Universit\'e Paris-Sud and CNRS UMR 8502, 91405 Orsay, France\\
$^{3}$Institut d'Electronique Fondamentale, Universit\'e Paris-Sud and CNRS UMR 8622, 91405 Orsay, France\\
$^{4}$ INAC-SPINTEC, Universit\'e Grenoble Alpes, CNRS and CEA, 38000 Grenoble, France\\
$^{5}$ INAC, CEA and Universit\'e Grenoble Alpes, 38054 Grenoble, France\\
$^{6}$ Singulus Technology AG, Hanauer Landstrasse 103, 63796 Kahl am Main, Germany}
\altaffiliation{These authors contributed equally to this work.}
\maketitle

{\bf
The recent observation of current-induced domain wall (DW) motion with large velocity in ultrathin magnetic wires has opened new opportunities for spintronic devices~\cite{Miron2011}. However, there is still no consensus on the underlying mechanisms of DW motion~\cite{Miron2011,Ryu2013,Emori2013,Haazen2013,Kim2013,Garello2013}. Key to this debate is the DW structure, which can be of Bloch or N\'eel type, and dramatically affects the efficiency of the different proposed mechanisms~\cite{Thiaville2012,Martinez2013,Brataas2014}. To date, most experiments aiming to address this question have relied on deducing the DW structure and chirality from its motion under additional in-plane applied fields, which is indirect and involves strong assumptions on its dynamics~\cite{Ryu2013,Emori2013,Haazen2013,Je2013}. Here we introduce a general method enabling direct, {\it in situ}, determination of the DW structure in ultrathin ferromagnets. It relies on local measurements of the stray field distribution above the DW using a scanning nanomagnetometer based on the Nitrogen-Vacancy defect in diamond~\cite{Taylor2008,Gopi2008,Rondin2014}.  We first apply the method to a Ta/Co$_{40}$Fe$_{40}$B$_{20}$(1~nm)/MgO magnetic wire and find clear signature of pure Bloch DWs. In contrast, we observe left-handed N\'eel DWs in a Pt/Co(0.6~nm)/AlO$_x$ wire, providing direct evidence for the presence of a sizable Dzyaloshinskii-Moriya interaction (DMI) at the Pt/Co interface. This method offers a new path for exploring interfacial DMI in ultrathin ferromagnets and elucidating the physics of DW motion under current.}

In wide ultrathin wires with perpendicular magnetic anisotropy (PMA), magnetostatics predicts that the Bloch DW, a helical rotation of the magnetization, is the most stable DW configuration because it minimizes volume magnetic charges~\cite{Hubert1998}. However, the unexpectedly large velocities of current-driven DW motion recently observed in ultrathin ferromagnets~\cite{Miron2011}, added to the fact that the motion can be found against the electron flow~\cite{Ryu2013,Emori2013}, has cast doubt on this hypothesis and triggered a major academic debate regarding the underlying mechanism of DW motion~\cite{Haazen2013,Kim2013,Garello2013,Thiaville2012,Martinez2013,Brataas2014}. Notably, it was recently proposed that N\'eel DWs with fixed chirality could be stabilized by the Dzyaloshinskii-Moriya interaction (DMI)~\cite{Thiaville2012}, an indirect exchange possibly occurring at the interface between a magnetic layer and a heavy metal substrate with large spin-orbit coupling~\cite{Crepieux1998}. For such chiral DWs, hereafter termed Dzyaloshinskii DWs, a damping-like torque due to spin-orbit terms (spin-Hall effect and Rashba interaction) would lead to efficient current-induced DW motion along a direction fixed by the chirality~\cite{Thiaville2012}. In order to validate unambiguously these theoretical predictions, a direct, {\it in situ}, determination of the DW structure in ultrathin ferromagnets is required. However, the relatively small number of spins at the wall center makes direct imaging of its inner structure a very challenging task. So far, only spin-polarized scanning tunnelling microscopy~\cite{Meckler2009} and spin-polarized low energy electron microscopy~\cite{Chen2013} have allowed a direct determination of the DW structure, demonstrating homochiral N\'eel DWs in Fe double layer on W(110) and in (Co/Ni)$_n$ multilayers on Pt or Ir, respectively. However, these techniques are intrinsically limited to model samples due to high experimental constraints and the debate remains open for widely used trilayer systems with PMA such as Pt/Co/AlO$_x$~\cite{Miron2011}, Pt/Co/Pt~\cite{Haazen2013} or Ta/CoFeB/MgO~\cite{Torrejon2013}. 

\begin{figure}[t]
\begin{center}
\includegraphics[width=0.65\textwidth]{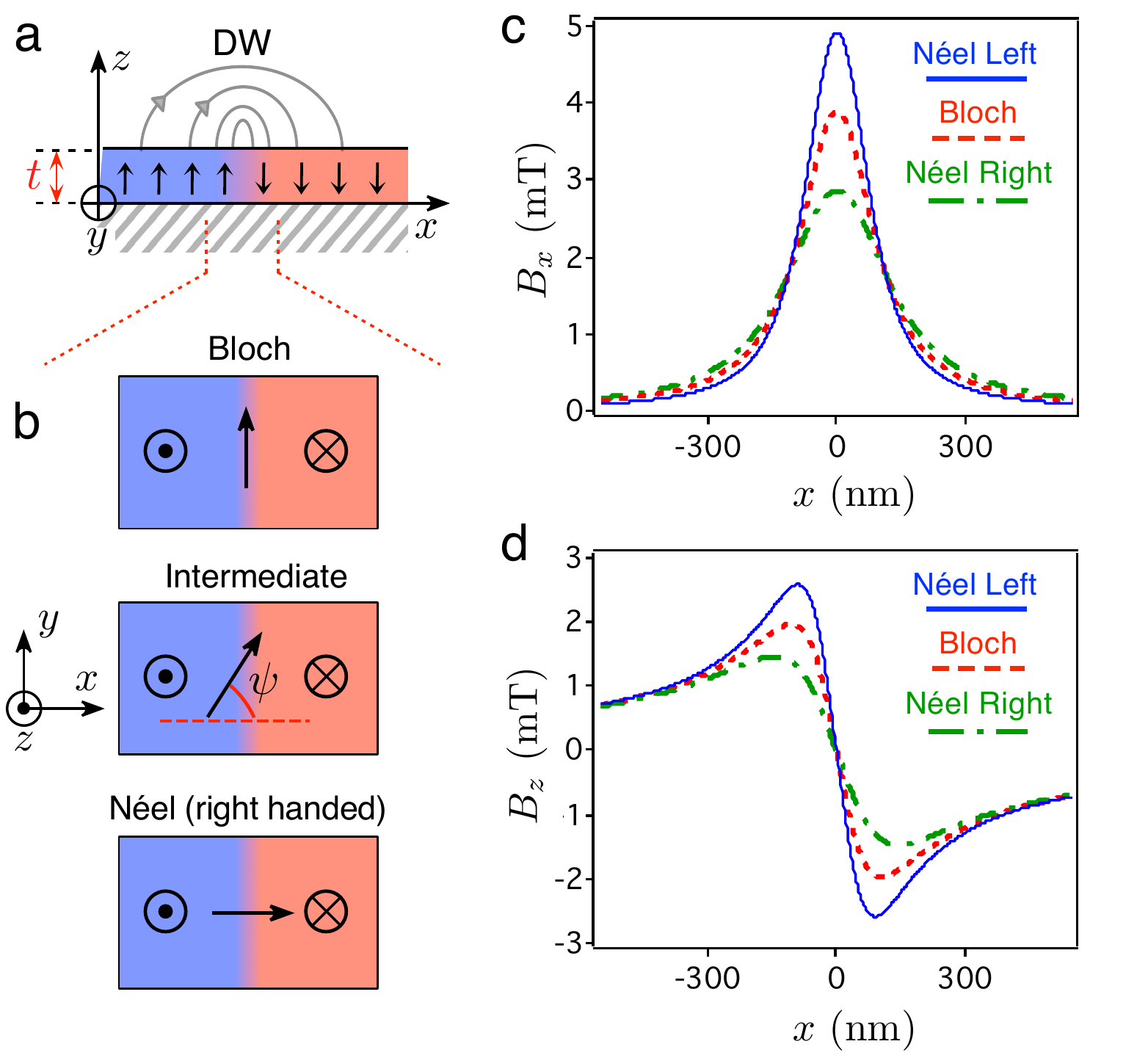}
\caption{{\bf Determining the nature of a DW by scanning nanomagnetometry.} {\bf a.} Schematic side view of a DW in a perpendicularly magnetized film. The black arrows indicate the internal magnetization while the grey arrows represent the magnetic field lines generated above the film. {\bf b.} Top view of the DW structure in a left-handed Bloch (top panel) or right-handed N\'eel (bottom panel) configuration, or an intermediate case characterized by the angle $\psi$. {\bf c,d.} Calculated stray field components $B_x^\psi(x)$ ({\bf c}) and $B_z^\psi(x)$ ({\bf d}) at a distance $d=120$ nm above the magnetic layer, with a DW centered at $x=0$. Here we use $M_s=10^6$ A/m and $\Delta_{\rm DW}=20$ nm.}
\label{Fig1}
\end{center}
\end{figure}

Here we introduce a general method which enables determining the nature of a DW in virtually any ultrathin ferromagnet. It relies on local measurements of the stray magnetic field produced above the DW using a scanning nanomagnetometer. To convey the basic idea behind our method, we start by deriving analytical formulas of the magnetic field distribution at a distance $d$ above a DW placed at $x=0$ in a perpendicularly magnetized film [Fig.~\ref{Fig1}a]. The main contribution to the stray field, denoted ${\bf B}^\perp$, is provided by the abrupt variation of the out-of-plane magnetization $M_z(x)=-M_s \tanh(x/\Delta_{\rm DW})$~\cite{Hubert1998}, where $M_s$ is the saturation magnetization and $\Delta_{\rm DW}$ is the DW width parameter. The resulting stray field components can be expressed as
\begin{equation}
\begin{dcases} 
B_x^\perp(x)\approx\frac{\mu_0M_st}{\pi}\frac{d}{x^2+d^2} \\
B_z^\perp(x)\approx-\frac{\mu_0M_st}{\pi}\frac{x}{x^2+d^2} \ ,
\end{dcases}
\label{eq:B_Bloch} 
\end{equation} 
where $t$ is the film thickness. These approximate formulas are valid in the limit of (i) $t\ll d$, (ii) $\Delta_{\rm DW}\ll d$ and (iii) for an infinitely long DW along the $y$ axis. On the other hand, the in-plane magnetization, with amplitude $M_\parallel(x)=M_s/\cosh(x/\Delta_{\rm DW})$, can be oriented with an angle $\psi$ with respect to the $x$ axis [Fig.~\ref{Fig1}b]. This angle is linked to the nature of the DW: $\psi=\pm \pi/2$ for a Bloch DW, whereas $\psi=0$ or $\pi$ for a N\'eel DW. The two possible values define the chirality (right or left) of the DW. The spatial variation of the in-plane magnetization adds a contribution ${\bf B}^{\parallel}\cos\psi $ to the stray field, whose components are given by
\begin{equation}
\begin{dcases} 
B_x^\parallel(x)\approx\frac{1}{2}\mu_0M_st\Delta_{\rm DW}\frac{x^2-d^2}{(x^2+d^2)^2} \\
B_z^\parallel(x)\approx\mu_0M_st\Delta_{\rm DW}\frac{xd}{(x^2+d^2)^2} \ .
\end{dcases}
\label{eq:B_IP} 
\end{equation}     
 
The net stray field above the DW is finally expressed as
\begin{equation}
{\bf B}^\psi(x)={\bf B}^\perp(x)+{\bf B}^\parallel(x)  \cos\psi \ ,
\end{equation}
which indicates that a N\'eel DW ($\cos\psi=\pm1$) produces an additional stray field owing to extra magnetic charges on each side of the wall. Using Eqs. (1) and (2), we find a maximum relative difference in stray field between Bloch and N\'eel DWs scaling as $\approx\pi\Delta_{\rm DW}/2d$. Local measurements of the stray field above a DW can therefore reveal its inner structure, characterized by the angle $\psi$. This is further illustrated in Figs.~\ref{Fig1}(c,d), which show the stray field components $B_x^\psi(x)$ and $B_z^\psi(x)$ for various DW configurations while using $d=120$~nm and $\Delta_{\rm DW}=20$~nm, which are typical parameters of the experiments considered below on a Ta/CoFeB(1nm)/MgO trilayer system.\\

\begin{figure}[t]
\begin{center}
\includegraphics[width=0.95\textwidth]{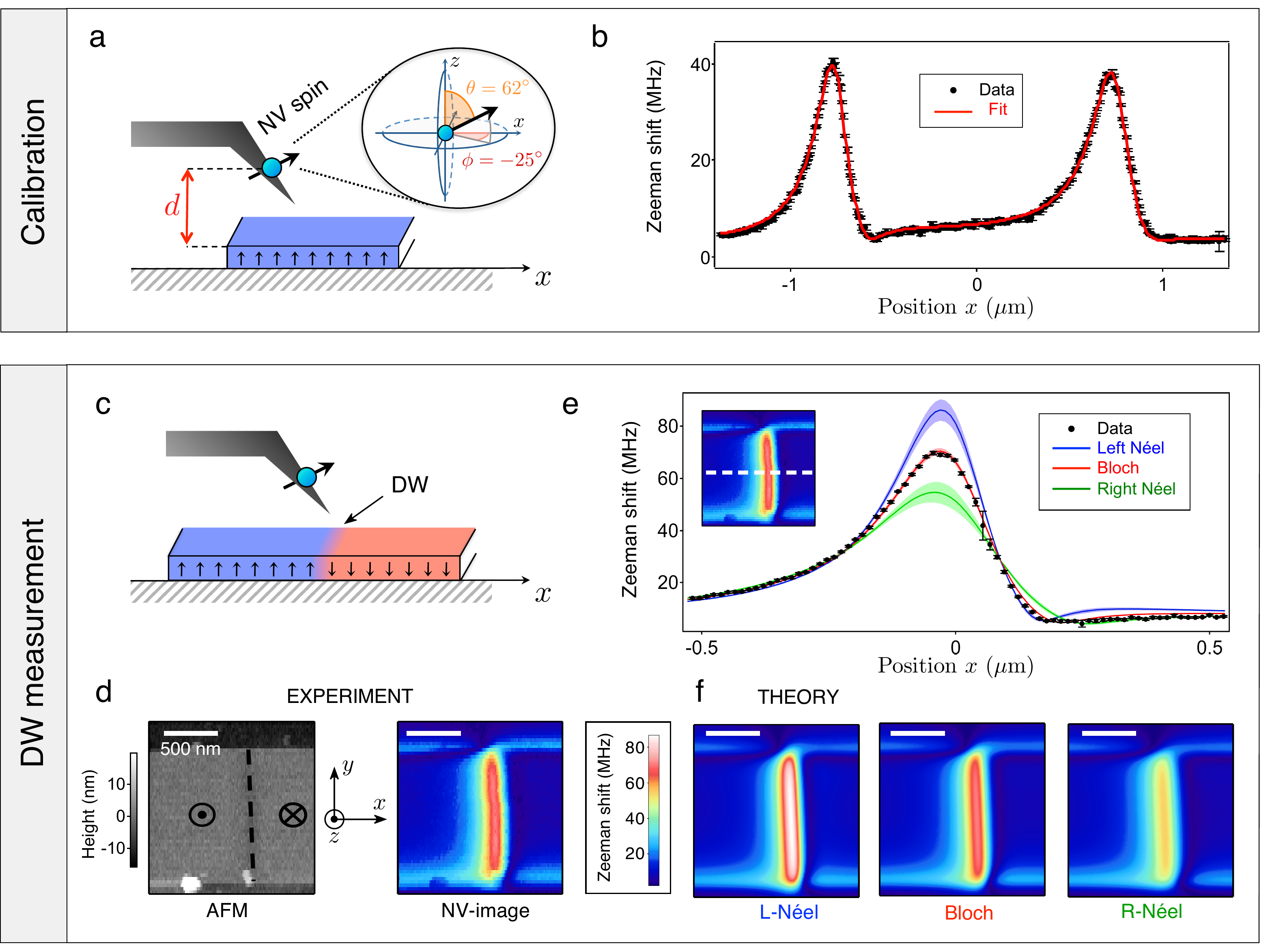}
\caption{{\bf Observation of Bloch DWs in a Ta/CoFeB/MgO wire.}
{\bf a.} The unknown parameters (distance $d$ and product $I_s=M_st$) are first calibrated by recording the stray field above a uniformly magnetized stripe. The inset defines the spherical angles ($\theta$,$\phi$) characterizing the NV spin's quantization axis, measured independently. {\bf b.} Zeeman shift of the NV spin measured as a function of $x$, across a 1.5-$\mu$m-wide stripe of Ta/CoFeB(1~nm)/MgO. The data (markers) are fitted to the theory (solid line), yielding $d=123\pm3$ nm and $I_s=926\pm26~\mu$A. {\bf c.} The stray field above a DW is then measured under the same conditions (same distance $d$, same NV spin). {\bf d.} AFM image (left panel) and corresponding Zeeman shift map (right panel) recorded on a 1.5-$\mu$m-wide stripe comprising a single DW. {\bf e.} Linecut across the DW (see dashed line in the inset). The markers are the experimental data, while the solid lines are the theoretical predictions for a Bloch (red), a left-handed N\'eel (blue) and a right-handed N\'eel DW (green). The shaded areas show 1 standard error in the simulations due to uncertainties in the parameters (Supplementary Section III-C). {\bf f.} Theoretical two-dimensional Zeeman shift maps for the same three DW configurations. In both {\bf e} and {\bf f}, the Bloch hypothesis is the one that best reproduces the data.}
\label{Fig2}
\end{center}
\end{figure}

We now demonstrate the effectiveness of the method by employing a single Nitrogen-Vacancy (NV) defect hosted in a diamond nanocrystal as a nanomagnetometer operating under ambient conditions~\cite{Taylor2008,Gopi2008,Rondin2014}. Here, the local magnetic field is evaluated within an atomic-size detection volume by monitoring the Zeeman shift of the NV defect electron spin sublevels through optical detection of the magnetic resonance. After grafting the diamond nanocrystal onto the tip of an atomic force microscope (AFM), we obtain a scanning nanomagnetometer which provides quantitative maps of the stray field emanating from nanostructured samples~\cite{Rondin2013,Tetienne2013,Tetienne2014} with a magnetic field sensitivity in the range of $10 \ \mu$T.Hz$^{-1/2}$~\cite{Rondin2012}. In this study, the Zeeman frequency shift $\Delta f_{\rm NV}$ of the NV spin is measured while scanning the AFM tip in tapping mode, so that the mean distance between the NV spin and the sample surface is kept constant with a typical tip oscillation amplitude of a few nanometers~\cite{Tetienne2013}. Each recorded value of $\Delta f_{\rm NV}$ is a function of $B_{\rm NV,\parallel}$ and $B_{\rm NV,\perp}$, which are the parallel and perpendicular components, respectively, of the local magnetic field with respect to the NV spin's quantization axis (Supplementary Section I). Note that a frequently found approximation is $\Delta f_{\rm NV}\approx g\mu_B B_{\rm NV,\parallel}/h$, where $g\mu_B/h\approx28$ GHz/T. This indicates that scanning-NV magnetometry essentially measures the projection $B_{\rm NV,\parallel}$ of the magnetic field along the NV center's axis. The latter is characterized by spherical angles ($\theta$,$\phi$), measured independently in the ($xyz$) reference frame of the sample [Fig.~\ref{Fig2}a]. \\

\begin{figure}[t]
\begin{center}
\includegraphics[width=0.8\textwidth]{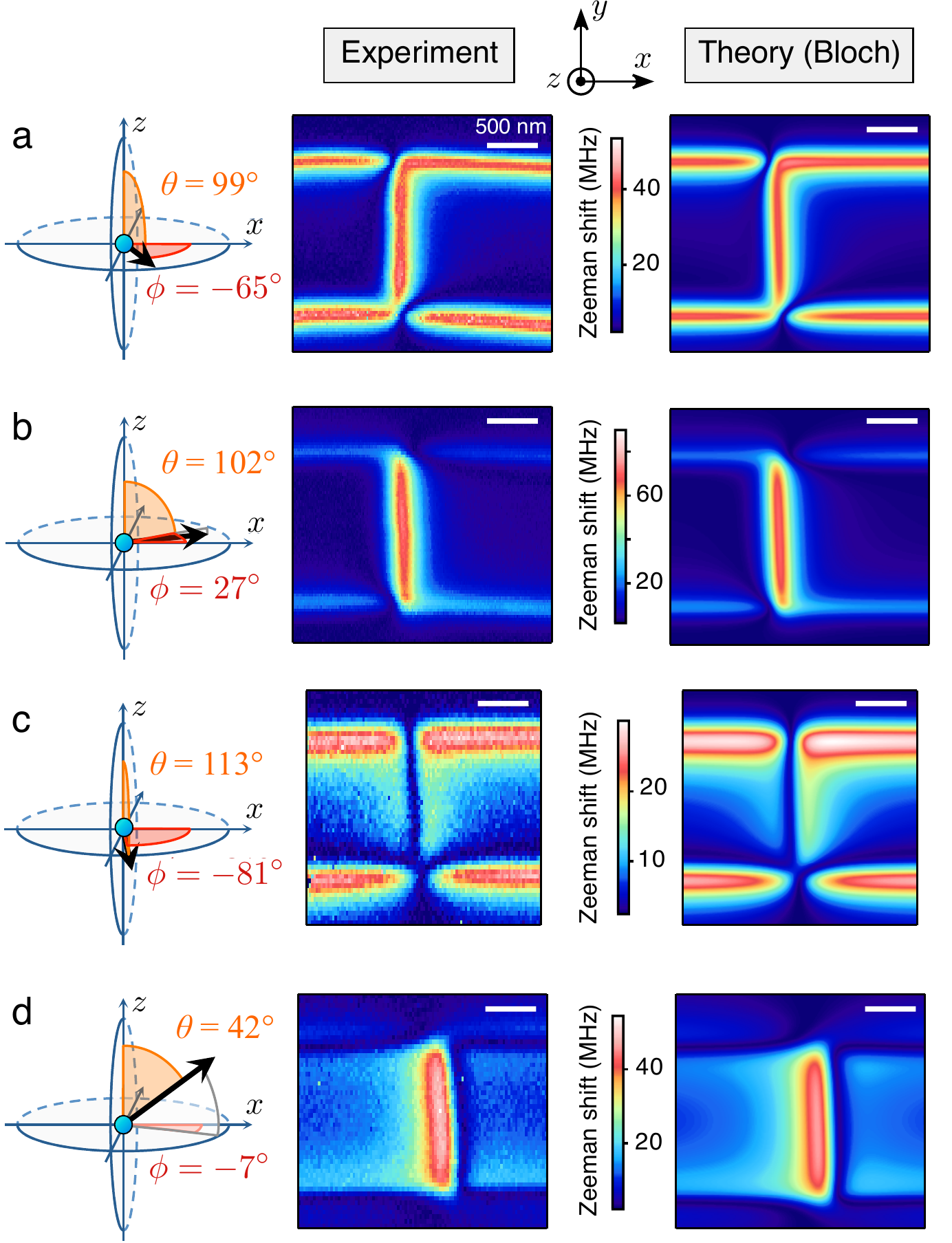}
\caption{
{\bf Vector mapping of the DW stray field.}
Zeeman shift maps above a DW in a Ta/CoFeB/MgO wire recorded with different magnetometer's projection axes (middle panels). The right panels shows the corresponding calculated maps assuming a Bloch-type DW. The distance $d$ is 116 nm ({\bf a}), 122 nm ({\bf b}), 195 nm ({\bf c}), and 178 nm ({\bf d}), respectively. The spherical angles ($\theta$,$\phi$) characterizing the projection axis are indicated on the left panels.}
\label{Fig3}
\end{center}
\end{figure}

\indent We first applied our method to determine the structure of DWs in a 1.5-$\mu$m-wide magnetic wire made of a Ta(5 nm)/Co$_{40}$Fe$_{40}$B$_{20}$(1 nm)/MgO(2 nm) trilayer stack (Supplementary Section II). This system has been intensively studied in the last years owing to low damping parameter and depinning field~\cite{Burrowes2013}. Before imaging a DW, it is first necessary to determine precisely (i) the distance $d$ between the NV probe and the magnetic layer and (ii) the product $I_s=M_st$, which are both directly involved in Eq.~(3). These parameters are obtained by performing a calibration measurement above the edges of an uniformly magnetized wire, as shown in Fig.~\ref{Fig2}a. Here we use the fact that the stray field profile ${\bf B}^{\rm edge}(x)$ above an edge placed at $x=0$ can be easily expressed analytically in a form similar to Eq.~(\ref{eq:B_Bloch}), which only depends on $d$ and $I_s$. An example of a measurement obtained by scanning the magnetometer across a Ta/CoFeB/MgO stripe is shown in Fig.~\ref{Fig2}b. The data are fitted with a function corresponding to the Zeeman shift induced by the stray field  ${\bf B}^{\rm edge}(x)-{\bf B}^{\rm edge}(x+w_c)$, where $w_c$ is the width of the stripe (Supplementary Section III-A). Repeating this procedure for a set of independent calibration linecuts, we obtain $d=123\pm3$ nm and $I_s=926\pm26~\mu$A, in good agreement with the value measured by other methods~\cite{Vernier2013}. \\
\indent Having determined all needed parameters, it is now possible to measure the stray field above a DW [Fig.~\ref{Fig2}c] and compare it to the theoretical prediction, which only depends on the angle $\psi$ that characterizes the DW structure. To this end, an isolated DW was nucleated in a wire of the same Ta/CoFeB/MgO film and imaged with the scanning-NV magnetometer under the same conditions as for the calibration measurements. The resulting distribution of the Zeeman shift $\Delta f_{\rm NV}$ is shown in Fig.~\ref{Fig2}d together with the AFM image of the magnetic wire. Within the resolving power of our instrument, limited by the probe-to-sample distance $d\sim 120$ nm~\cite{Tetienne2013}, the DW appears to be straight with a small tilt angle with respect to the wire long axis, determined to be $2\pm 1^\circ$ (Supplementary Section III-B). Taking into account this DW spatial profile, the stray field above the DW was computed for (i) $\psi=0$ (right-handed N\'eel DW), (ii) $\psi=\pi$ (left-handed N\'eel DW) and (iii) $\psi\pm\pi/2$ (Bloch DW). Here we used the micromagnetic OOMMF software~\cite{oommf,Rohart2013} rather than the analytical formula described above in order to avoid any approximation in the calculation. The computed magnetic field distributions were finally converted into Zeeman shift distribution taking into account the NV spin's quantization axis. A linecut of the experimental data across the DW is shown in Fig.~\ref{Fig2}e, together with the predicted curves in the three above-mentioned cases. Excellent agreement is found if one assumes that the DW is purely of Bloch type. The same conclusion can be drawn by directly comparing the full two-dimensional theoretical maps to the data [Fig.~\ref{Fig2}d and f]. As described in detail in the Supplementary Section III-C, all sources of uncertainty in the theoretical predictions were carefully analysed, yielding the 1 standard error (s.e.) intervals shown as shaded areas in Fig.~\ref{Fig2}e. Based on this analysis, we find a 1 s.e. upper limit $|\cos\psi|<0.07$. This corresponds to an upper limit for the DMI parameter $D_{\rm DMI}$, as defined in Ref.~\cite{Thiaville2012}, of $|D_{\rm DMI}|<0.01$ mJ/m$^2$ (Supplementary Section III-C). This result was confirmed on a second DW in the same wire. In addition, the measurements were reproduced for different projection axes of the NV probe. The results are shown in Fig.~\ref{Fig3} for four NV defects with different quantization axes, showing excellent agreement between experiment and theory if one assumes a Bloch-type DW. These experiments provide an unambiguous confirmation of the Bloch nature of the DWs in our sample, but are also a striking illustration of the vector mapping capability offered by NV microscopy, allowing for robust tests of theoretical predictions. \\
\indent We conclude that there is no evidence for the presence of a sizable interfacial DMI in a Ta(5nm)/Co$_{40}$Fe$_{40}$B$_{20}$(1nm)/MgO trilayer stack. This is in contrast with recent experiments reported on similar samples with different compositions, such as Ta(5nm)/Co$_{80}$Fe$_{20}$(0.6nm)/MgO~\cite{Emori2013,Emori2013b} and Ta(0.5 nm)/Co$_{20}$Fe$_{60}$B$_{20}$(1nm)/MgO~\cite{Torrejon2013}, where indirect evidence for N\'eel DWs was found through current-induced DW motion experiments. We note that contrary to these studies, our method indicates the nature of the DW at rest, in a direct manner, without any assumption on the DW dynamics. Our results therefore motivate a systematic study of the DW structure upon modifications of the composition of the trilayer stack. \\

\begin{figure}[t]
\begin{center}
\includegraphics[width=0.77\textwidth]{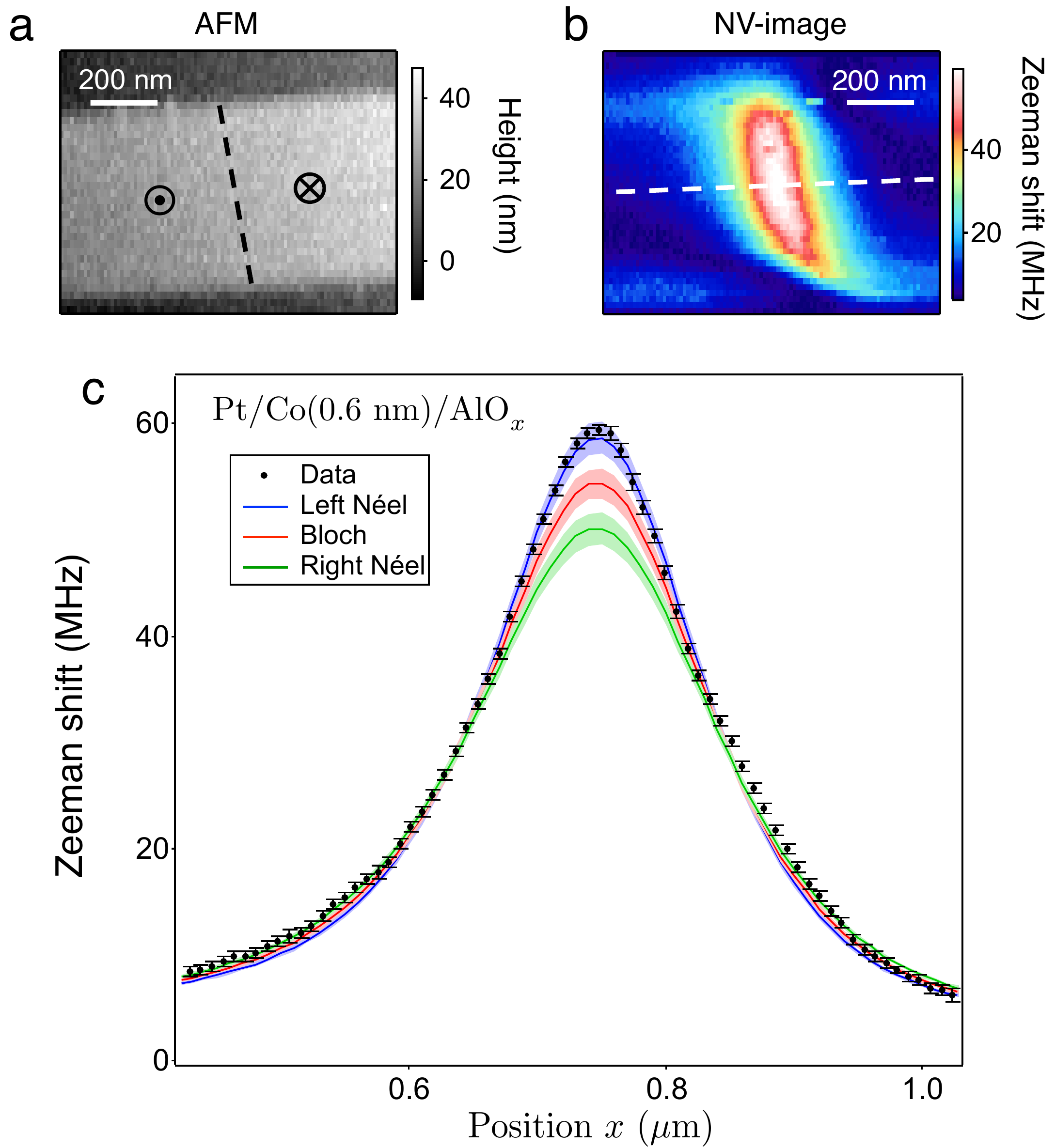}
\caption{{\bf Observation of left-handed N\'eel DWs in a Pt/Co/AlO$_x$ wire.}
{\bf (a)} AFM image and {\bf (b)} corresponding Zeeman shift map recorded by scanning the NV magnetometer above a DW in a 500-nm-wide magnetic wire of Pt/Co(0.6~nm)/AlO$_x$. The spherical angles characterizing the NV defect quantization axis are $(\theta=87^{\circ},\phi=23^{\circ})$. From calibration measurements above edges of the sample, we inferred $d=119.0\pm3.4$ nm and $I_s=671\pm18 \ \mu$A (Supplementary Section III-A). {\bf c.} Linecut extracted from {\bf b} (markers), together with the theoretical prediction (solid lines) for a Bloch (red), a left-handed N\'eel (blue) and a right-handed N\'eel DW (green). The shaded areas indicates 1 s.e. uncertainty in the simulations (Supplementary Section III-C). }
\label{Fig4}
\end{center}
\end{figure}

In a second step, we explored another type of sample, namely a Pt(3nm)/Co(0.6 nm)/AlO$_x$(2nm) trilayer grown by sputtering on a thermally oxidized silicon wafer (Supplementary Section II). The observation of current-induced DW motion with unexpectedly large velocities in this asymmetric stack has attracted considerable interest in the recent years~\cite{Miron2011}. Here, the DW width is $\Delta_{\rm DW}\approx 6$ nm, leading to a relative field difference between Bloch and N\'eel cases of $\approx 8\%$ at a distance $d\approx 120$~nm. We followed a procedure similar to that described above (Supplementary Section III). After a preliminary calibration of the experiment, a DW in a 500-nm-wide magnetic wire was imaged [Fig.~\ref{Fig4}a,b] and linecuts across the DW were compared to theoretical predictions [Fig.~\ref{Fig4}c]. Here the experimental results clearly indicate a N\'eel-type DW structure with left-handed chirality. The same result was found for two other DWs. This provides direct evidence of a strong DMI at the Pt/Co interface, with a lower bound $|D_{\rm DMI}|>0.1$ mJ/m$^2$. This result is consistent with the conclusions of recent field-dependent DW nucleation experiments performed in similar films~\cite{Pizzini2014}. In addition, we note that the observed left-handed chirality, once combined with a damping-like torque induced by the spin-orbit terms, could explain the characteristics of DW motion under current in this sample~\cite{Martinez2013}. 

In conclusion, we have shown how scanning-NV magnetometry enables direct discrimination between competing DW configurations in ultrathin ferromagnets. This method, which is not sensitive to possible artifacts linked to the DW dynamics, will help clarifying the physics of DW motion under current, a necessary step towards the development of DW-based spintronic devices. In addition, this work opens a new avenue for studying the mechanisms at the origin of interfacial DMI in ultrathin ferromagnets, by measuring the DW structure while tuning the properties of the magnetic material~\cite{Torrejon2013,Ryu2014}. This is a key milestone in the search for systems with large DMI that could sustain magnetic skyrmions~\cite{Fert2013}.\\

\noindent {\bf Aknowledgements}. This research has been partially funded by the European Community's Seventh Framework Programme (FP7/2007-2013) under Grant Agreement n$^{\circ}$ 611143 (D{\sc iadems}) and n$^{\circ}$ 257707 (M{\sc agwire}), the French Agence Nationale de la Recherche through the projects D{\sc iamag} and E{\sc sperado}, and by C'Nano Ile-de-France (N{\sc anomag}).\\

\noindent {\bf Author contributions}. S.R. and A.T. conceived the idea of the study. J.P.T., T.H., L.J.M. and V.J. performed the experiments, analysed the data and wrote the manuscript. L.H.D, K.G., J.P.A., G.G., L.V., and B.O. prepared the samples. All authors discussed the data and commented on the manuscript.
\newpage

\newpage

\newpage

\newpage

\newpage
\newpage
\clearpage
\begin{center}
{\Large Supplementary Informations}
\end{center}

\section{Scanning-NV magnetometry}

The experimental setup combines a tuning-fork-based atomic force microscope (AFM) and a confocal optical microscope (attoAFM/CFM, Attocube Systems), all operating under ambient conditions. A detailed description of the setup as well as the method to graft a diamond nanocrystal onto the apex of the AFM tip can be found in Ref. [\onlinecite{Rondin2012}]. 

\subsection{Characterization of the magnetic field sensor} \label{NVcharac}

The data reported in this work were obtained with NV center magnetometers hosted in three different nanodiamonds, labeled ND74 (data of Figure 3 of the main paper), ND75 (Figure 2) and ND79 (Figure 4). All nanodiamonds were $\approx 50$ nm in size, as measured by AFM before grafting the nanodiamond onto the AFM tip. The magnetic field was inferred by measuring the Zeeman shift of the electron spin resonance (ESR) of the NV center's ground state~[\onlinecite{Rondin2014}]. This is achieved by monitoring the spin-dependent photoluminescence (PL) intensity of the NV defect while sweeping the frequency of a CW radiofrequency (RF) field generated by an antenna fabricated directly on the sample.

\begin{figure}[t]
\begin{center}
\includegraphics[width=0.65\textwidth]{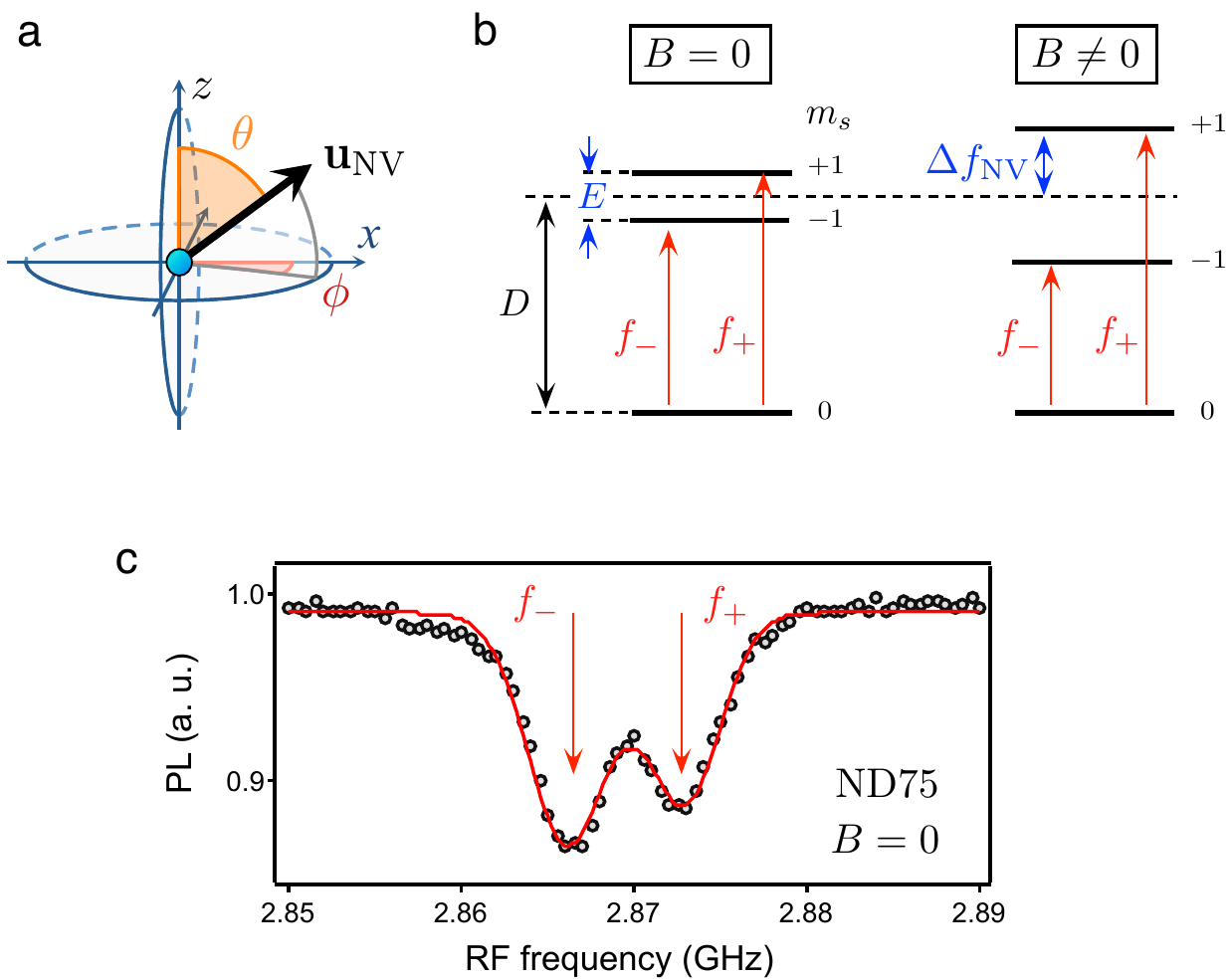}
\caption{(a) The quantization axis ${\bf u}_{\rm NV}$ of the NV center's electron spin is characterized by spherical angles ($\theta$,$\phi$) in the $(xyz)$ sample reference frame. (b) Structure of the spin sublevels in the NV defect's ground state. The ESR frequencies corresponding to the electron spin transitions $m_s=0\rightarrow -1$ and $m_s=0\rightarrow +1$ are denoted by $f_-$ and $f_+$, respectively. (c) ESR spectrum of ND75 in zero external magnetic field. A fit to a sum of two Gaussian functions allows determining the parameters $D$ and $E$ that characterize the NV center. The values are given in Table \ref{tab1}.} 
\label{FigS_NVcharac}
\end{center}
\end{figure}

The Hamiltonian used to describe the magnetic-field dependence of the two ESR transitions of this $S=1$ spin system is given by 
\begin{equation} \label{Hamilto}
{\cal H}=hDS_{Z}^{2}+hE(S_X^{2}-S_Y^{2})+g\mu_B\mathbf{B}\cdot\mathbf{S} \ ,
\end{equation}
where $D$ and $E$ are the zero-field splitting parameters that characterize a given NV center, $h$ is the Planck constant, $g\mu_B/h=28.03(1)$ GHz/T~[\onlinecite{Doherty2012}], {\bf B} is the local magnetic field and {\bf S} is the dimensionless $S=1$ spin operator. Here, the $(XYZ)$ reference frame is defined by the diamond crystal orientation, with $Z$ being parallel to the NV center's symmetry axis ${\bf u}_{\rm NV}$, as shown in Figure~\ref{FigS_NVcharac}(a). 

The two ESR frequencies are denoted $f_+$ and $f_-$ and the Zeeman shifts are defined by $\Delta f_\pm=f_\pm-D$ [Fig. \ref{FigS_NVcharac}(b)]. In general, $\Delta f_\pm$ are functions of $B_{\rm NV,\parallel}=|{\bf B}\cdot {\bf u}_{\rm NV}|$ and $B_{\rm NV,\perp}=\Vert {\bf B}\times {\bf u}_{\rm NV} \Vert$. However, in the limit of small transverse fields ($g\mu_B B_{\rm NV,\perp} \ll hD$) [\onlinecite{Tetienne2012}], they depend only on the magnetic field projection along the NV axis $B_{\rm NV,\parallel}$, following the relation
\begin{equation} \label{eqESRfreq}
\Delta f_\pm(B_{\rm NV,\parallel})\approx \pm\sqrt{(g\mu_B B_{\rm NV,\parallel}/h)^2+E^{2}} \ .
\end{equation}  
The parameters $D$ and $E$ were extracted from ESR spectra recorded at zero magnetic field using the fact that $f_{\pm}({\bf B}={\bf 0})=D\pm E$ [see Fig. \ref{FigS_NVcharac}(b,c)]. In all the data shown in this work, only the upper frequency $f_+$ was measured. Thereafter, we will note the corresponding Zeeman shift $\Delta f_{\rm NV}=f_+-D$, the subscript `NV' reminding that it depends on the direction ${\bf u}_{\rm NV}$ [Fig. \ref{FigS_NVcharac}(b)]. The experimental measurements of $\Delta f_{\rm NV}$ were compared to theory by calculating the expected Zeeman shift through full diagonalization of the Hamiltonian defined by Eq.~(\ref{Hamilto}), given the theoretical {\bf B} map. However, we note that since the condition $g\mu_B B_{\rm NV,\perp} \ll hD$ is usually fulfilled in our measurements, the formula (\ref{eqESRfreq}) is approximately valid, so that in principle one could retrieve directly the value of $B_{\rm NV,\parallel}$ with good accuracy ($<0.1$ mT). 

The nanodiamonds were recycled several times to be used with different orientations ${\bf u}_{\rm NV}$ with respect to the $(xyz)$ reference frame of the sample. The various orientations are labeled with small letters: ND74a, ND74d, ND74e, ND74g, ND75c, ND79c. The spherical angles ($\theta$,$\phi$) that characterize the direction ${\bf u}_{\rm NV}$ were obtained by applying an external magnetic field of known direction and amplitude with a three-axis coil system, following the procedure described in Ref.~[\onlinecite{Rondin2013}]. The measurement uncertainty of $2^\circ$ (standard error) is related to the precision of the calibration of the coils and their alignment with respect to the $(xyz)$ reference frame. 

Table \ref{tab1} indicates the parameters $D$, $E$, $\theta$ and $\phi$ measured for each NV magnetometer used in this work, with the associated standard errors.\\

\begin{table}[htb!]
\begin{center}
\begin{tabular}{|c|c|c|c|c|c|c|}
\hline
 & ND74a & ND74d & ND74e & ND74g & ND75c & ND79c \\
\hline
\hline
 Figure & 3(a) & 3(b) & 3(c) & 3(d) & 2 & 4 \\
 \hline
$D$ ($\pm 0.2$ MHz) & \multicolumn{4}{c|}{2867.1} & 2869.5 & 2866.6 \\
\hline
$E$ ($\pm 0.2$ MHz) & \multicolumn{4}{c|}{3.1} & 3.3 & 4.3 \\
\hline
$\theta$ ($\pm 2^\circ$) & $99^{\circ}$ & $102^{\circ}$ & $113^{\circ}$ & $42^{\circ}$ &  $62^{\circ}$ & $87^{\circ}$ \\
\hline
$\phi$ ($\pm 2^\circ$) & $-65^{\circ}$ & $27^{\circ}$ & $-81^{\circ}$ & $-7^{\circ}$ &  $-25^{\circ}$ & $23^{\circ}$ \\
\hline
\end{tabular}
\caption{Summary of the parameters ($D,E,\theta,\phi$) measured for the different NV center magnetometers used in this work. The second row mentions the figures of the main paper where the magnetometer is used.}
\label{tab1}
\end{center}
\end{table}

\subsection{Quantitative stray field mapping} \label{ESRmapping}

\begin{figure}[t]
\begin{center}
\includegraphics[width=0.8\textwidth]{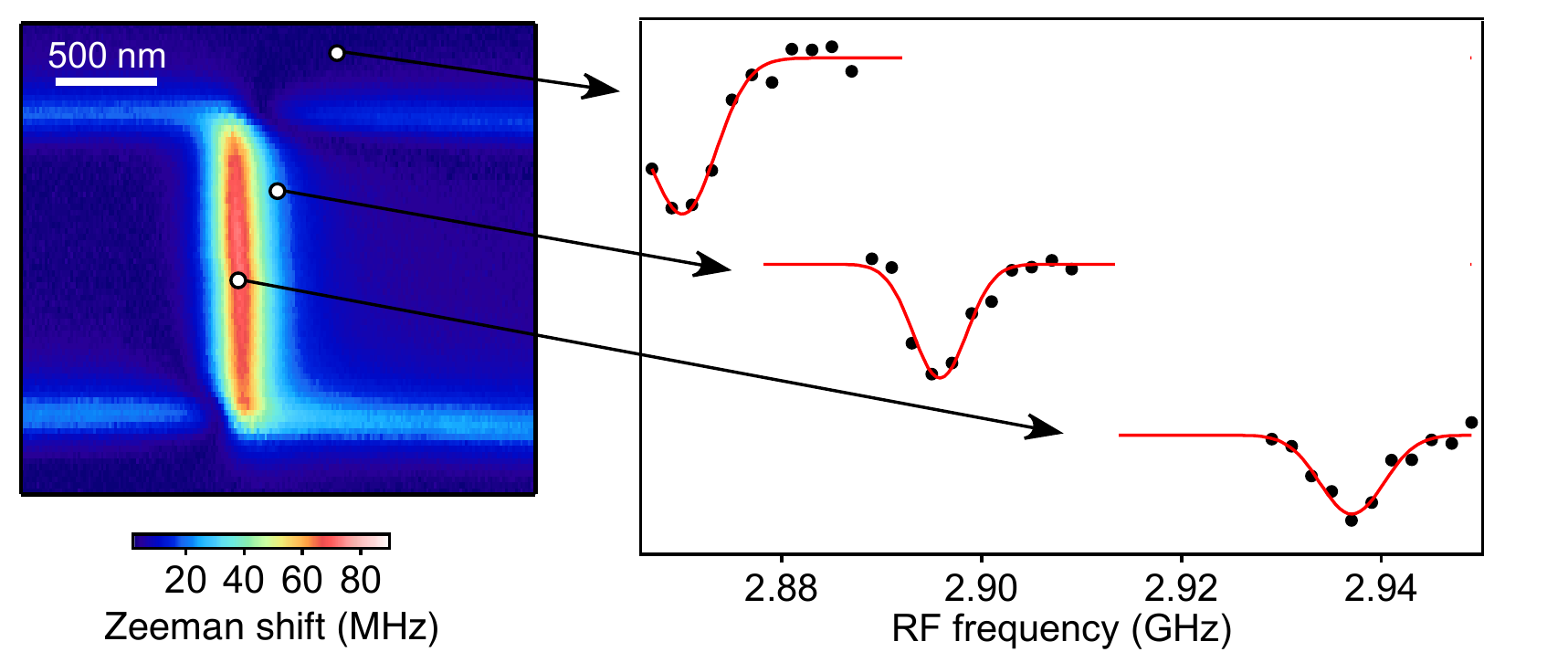}
\caption{Map of the Zeeman shift $\Delta f_{\rm NV}$ obtained with ND74d above the domain wall (reproduced from Fig. 3b of the main paper), along with the raw ESR spectra corresponding to 3 different selected pixels. Solid lines are Gaussian fits.} 
\label{FigS_ESRmapping}
\end{center}
\end{figure}

The experimental Zeeman shift maps were obtained by recording ESR spectra while scanning the magnetometer with the AFM operated in tapping mode. Each spectrum is composed of 11 bins with a bin size of 2 MHz, leading to a full range of 20 MHz. The integration time per bin is 40 ms, hence 440 ms per spectrum, that is, 440 ms per pixel of the image. As illustrated in Figure~\ref{FigS_ESRmapping}, only the upper frequency $f_+$ is probed, and the measurement window is shifted from pixel to pixel in order to track the resonance. Each spectrum is then fitted with a Gaussian lineshape to obtain $f_+$ and thus $\Delta f_{\rm NV}$. The full width at half maximum (FWHM) is typically 5-10 MHz, and the standard error on $f_+$ is $\approx 0.3$ MHz with the above-mentioned acquisition parameters.

\section{Samples} \label{samples}

Two samples, Ta/CoFeB/MgO and Pt/Co/AlO$_x$, were investigated in this work. The Ta/CoFeB/MgO trilayer was deposited on a Si/SiO$_2$(100 nm) wafer using a PVD Timaris deposition tool by Singulus Tech. The film stack composition is Ta(5)/CoFeB(1)/MgO(2)/Ta(5), starting from the SiO$_2$ layer (units in nanometer). The stoichiometric composition of the as-deposited magnetic layer is Co$_{40}$Fe$_{40}$B$_{20}$. The second sample was fabricated from Pt(3)/Co(0.6)/Al(1.6) layers deposited on a thermally oxidized silicon wafer by d.c. magnetron sputtering. After deposition, the aliminium layer was oxidized by exposure to an oxygen plasma. Both samples were patterned using e-beam lithography and ion milling. A second step of e-beam lithography was finally performed in order to define a gold stripline for RF excitation, which is used to record the Zeeman shift of the NV defect magnetometer [cf. section~\ref{NVcharac}]. 

\begin{figure}[t]
\begin{center}
\includegraphics[width=0.95\textwidth]{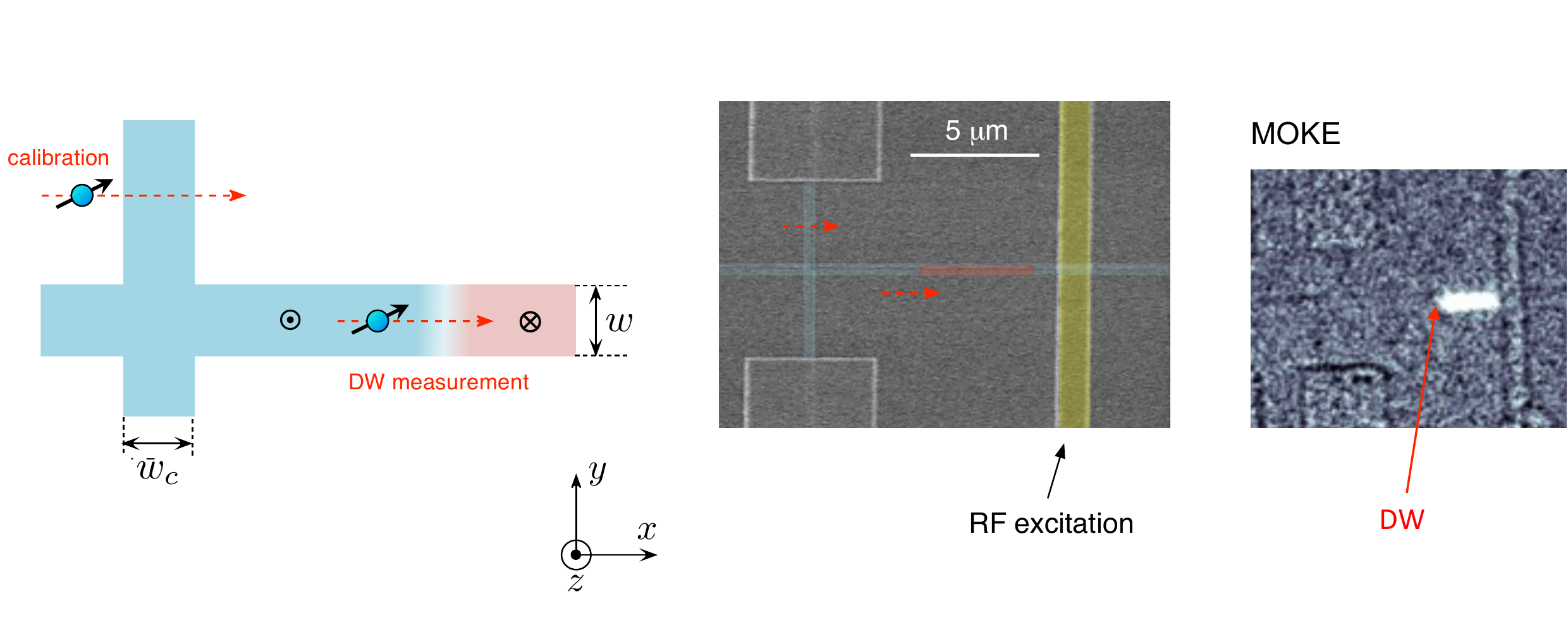}
\caption{The samples were patterned into two perpendicular wires, one of width $w_c$ used for the calibration, the other of width $w$ for the DW study. Left panel: Schematic of the sample. Middle panel: Scanning electron micrograph of the Ta/CoFeB/MgO sample, showing in color the magnetic domains (up in blue, down in red) and the RF antenna (yellow). Right panel: Magneto-optical Kerr microscopy image of the Ta/CoFeB/MgO sample after nucleation.}
\label{FigS_sample}
\end{center}
\end{figure}

Figure~\ref{FigS_sample} shows the general schematic of the samples, highlighting the regions used for calibration linecuts (stripe of width $w_c$) and DW measurements (stripe of width $w$) [cf. section~\ref{calib}]. The use of two perpendicular wires ensures that the DW is approximately parallel to the edges used for calibration. The final dimensions (height $\delta d_m$, widths $w_c$ and $w$) of the patterned structures were measured with a calibrated AFM. For the Ta/CoFeB/MgO sample, $\delta d_m=17\pm 2$ nm and $w_c=w=1500\pm30$ nm, whereas for the Pt/Co/AlO$_x$ sample $\delta d_m=25\pm 3$ nm, $w_c=980\pm20$ nm and $w=470\pm20$ nm. The nucleation was achieved by feeding a current pulse through the gold stripline for the Ta/CoFeB/MgO sample, and by applying pulses of out-of-plane magnetic field for the Pt/Co/AlO$_x$ sample.

\section{Data analysis} 

\subsection{Calibration linecuts} \label{calib}

\subsubsection{Fit procedure}

\begin{figure}[t]
\begin{center}
\includegraphics[width=1\textwidth]{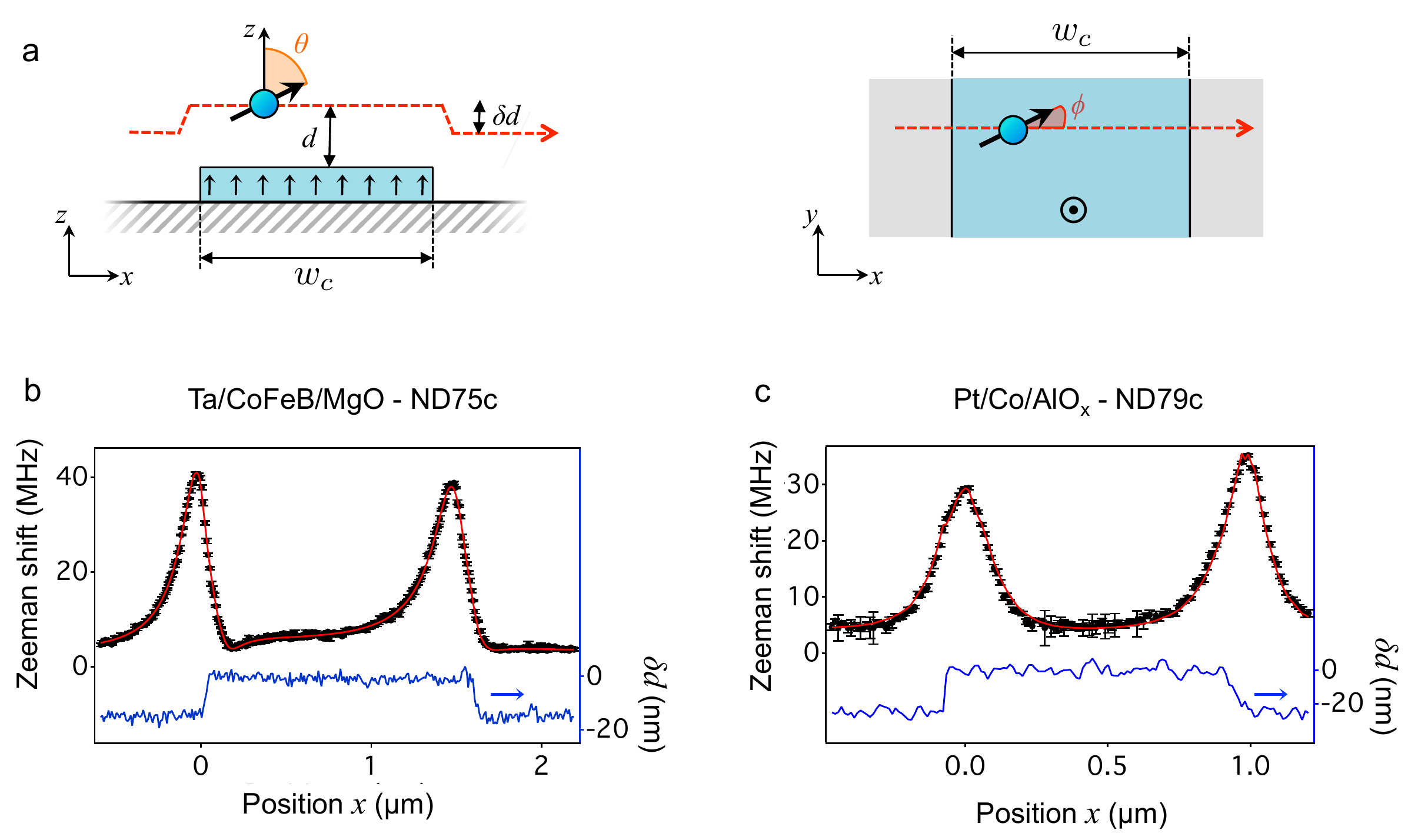}
\caption{(a) Principle of the calibration experiment. The Zeeman shift $\Delta f_{\rm NV}^{\rm stripe}(x)$ is measured while scanning the NV magnetometer across a stripe of the ferromagnetic layer in the $x$ direction. (b,c) Zeeman shift linecuts measured with ND75c across a stripe of Ta/CoFeB/MgO (b) and across a stripe of Pt/Co/AlO$_x$ with ND79c (c). The red solid line is the fit, as explained in the text. The blue curve is the topography of the sample recorded simultaneously by the AFM and used to define the distance change $\delta d(x)$ in the fit function.} 
\label{FigS_Linecuts}
\end{center}
\end{figure}

As discussed in the main text, a preliminary calibration of the experiment is required in order to infer the probe-sample distance $d$ and the saturation magnetization of the sample $M_s$. This calibration is performed by measuring the Zeeman shift $\Delta f_{\rm NV}$ of the NV magnetometer while scanning it across a stripe of the ferromagnetic layer in the $x$ direction, as depicted in Figure~\ref{FigS_Linecuts}(a). Since $d$ is of the order of 100 nm in our experiments, one has $d\gg t$ where $t$ is the thickness of the magnetic layer, so that the edges can be considered to be abrupt, {\it i.e.} $M_z(-w_c<x<0)=M_s$ and $M_z=0$ otherwise, with $w_c$ the stripe width. In fact, due to the topography of the sample, the effective distance between the NV spin and the magnetic layer varies during the scan [see Fig.~\ref{FigS_Linecuts}(a)]. This position-dependent distance can be written as $d_{\rm eff}(x)=d+\delta d(x)$, where $\delta d(x)=0$ on average when the tip is above the stripe, and $\delta d(x)=-\delta d_m$ on average when the tip is above the bare substrate. Here $\delta d_m$ is the total height of the patterned structures [cf. Section~\ref{samples}]. Experimentally, one has access to the relative variations of $d_{\rm eff}(x)$ thanks to the simultaneously recorded AFM topography information, hence one can infer the function $\delta d(x)$. Therefore, only the absolute distance, characterized by $d$, is unknown. 

The stray field components above a single abrupt edge parallel to the $y$ direction, positioned at $x=0$ (magnetization pointing upward for $x<0$), are given by  
\begin{equation}
\begin{dcases} 
B_x^{\rm edge}(x)=\frac{\mu_0M_st}{2\pi}\frac{d_{\rm eff}(x)}{x^2+d_{\rm eff}^2(x)} \\
B_y^{\rm edge}(x)=0 \\
B_z^{\rm edge}(x)=-\frac{\mu_0M_st}{2\pi}\frac{x}{x^2+d_{\rm eff}^2(x)}~.
\end{dcases}
\label{eq:Bedge} 
\end{equation} 
These formulas correspond to the thin-film limit ($d\gg t$) of exact formulas, but the relative error introduced by the approximation is $<10^{-5}$ in our case ($d/t \sim100$), which is negligible compared with other sources of error (see below). The field above a stripe is then obtained by simply adding the contribution of the two edges, namely 
\begin{equation} \label{eq:Bstripe} 
{\bf B}^{\rm stripe}(x)={\bf B}^{\rm edge}(x)-{\bf B}^{\rm edge}(x+w_c)~.
\end{equation} 
Using Eqs.~(\ref{eq:Bedge}) and (\ref{eq:Bstripe}), we obtain an analytical formula for the stray field above the stripe. A fit function $\Delta f_{\rm NV}^{\rm stripe}(x)$ is then obtained by converting the field distribution into Zeeman shift of the NV defect after diagonalization of the Hamiltonian defined by Eq.~(\ref{Hamilto}), with the characteristic parameters ($\theta,\phi,E,D$) of the NV magnetometer. The fit parameters are the maximum distance $d$ and the product $I_s=M_st$. The geometric parameters of the stripe (width $w_c$ and height $\delta d_m$), measured independently, serve as references to rescale the length scales $x$ and $z$ in the linecut data before fitting. Note that in assuming an uniformely magnetized stripe, we neglect the rotation of the magnetization near the edges induced by the Dzyaloshinskii-Moriya interaction (DMI)~[\onlinecite{Rohart2013}]. The effect of this rotation will be discussed in Section \ref{DMItilt}.

In the following, we focus on the experiments performed (i) with ND75c on the Ta/CoFeB/MgO sample and (ii) with ND79c on the Pt/Co/AlO$_x$ sample, corresponding to the experimental results reported in Figures 2 and 4 of the main paper, respectively. Typical calibration linecuts are shown in Figures~\ref{FigS_Linecuts}(b,c) together with the topography of the sample. The red solid line is the result of the fit, showing a very good agreement with the experimental data. 

\subsubsection{Uncertainties}

Uncertainties on the fit parameters $X=\{I_s,d\}$ come from those on (i) the NV center's parameters $(\theta,\phi,E,D)$, (ii) the geometric parameters of the stripe ($w_c,\delta d_m$) and (iii) the fit procedure. There are therefore six independent parameters $\{p_i\}=\{\theta,\phi,E,D,w_c,\delta d_m\}$ which introduce uncertainties on the outcome of the fit. In the following, these parameters are denoted as $p_i=\bar{p_{i}} \pm \sigma_{p_{i}}$ where $\bar{p_{i}}$ is the nominal value of  parameter $p_i$ and $\sigma_{p_{i}}$ its standard error. The uncertainties on $\theta$, $\phi$, $E$ and $D$ (resp. on $w_c$ and $\delta d_m$) are discussed in Section~\ref{NVcharac} (resp. in Section \ref{samples}). The nominal values and the standard errors on each parameter $p_i$ are summarized in Table~\ref{tab0}.

\begin{figure}[t]
\begin{center}
\includegraphics[width=0.88\textwidth]{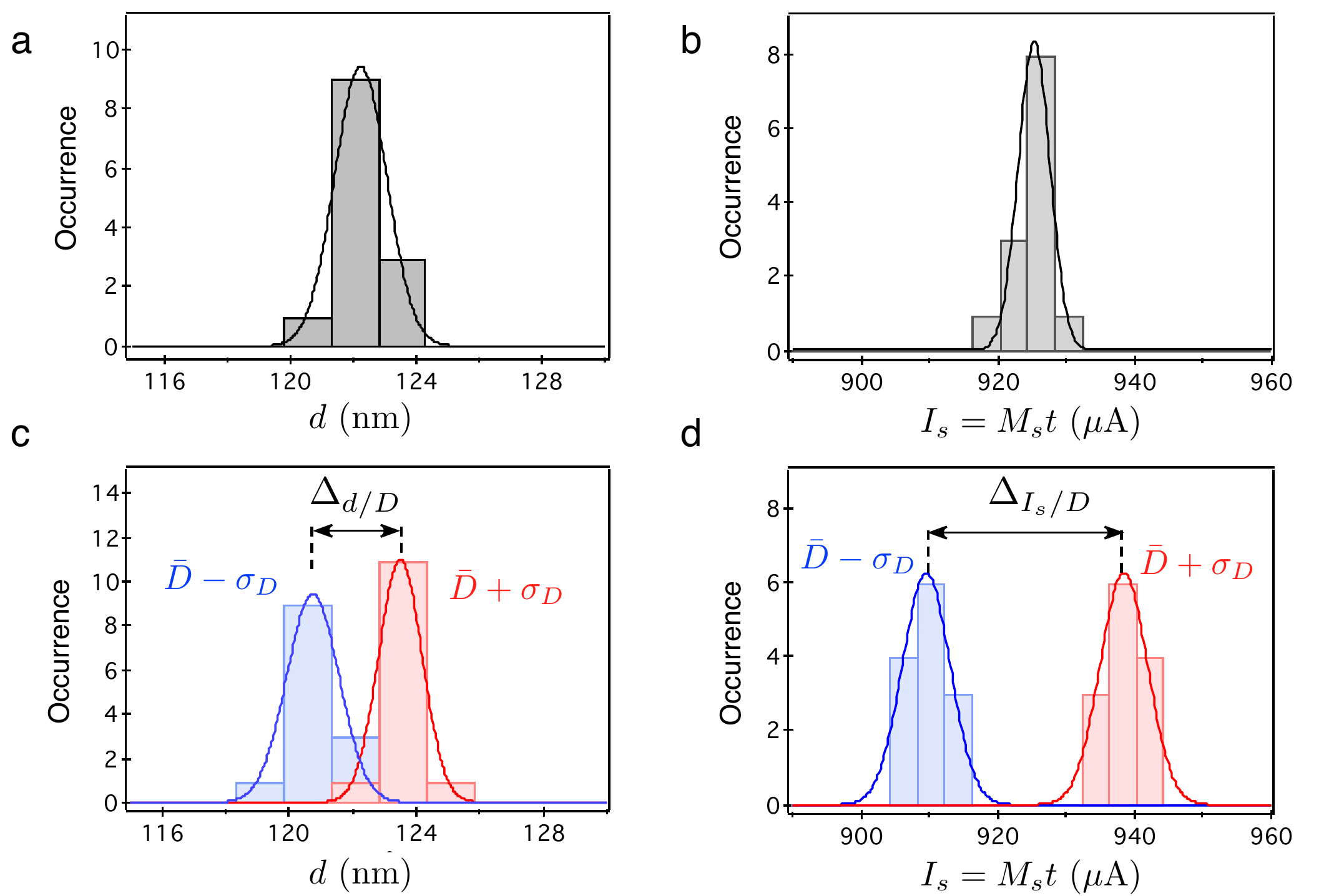}
\caption{(a,b) Histograms of the fit outcomes for the probe-sample distance $d$ (a) and $I_s=M_st$ (b) obtained for a set of 13 calibration linecuts on the Ta/CoFeB/MgO sample with ND75c while fixing the parameters $p_i$ to their nominal values $\bar{p_{i}}$. (c,d) Histograms of the fit outcomes with the zero-field splitting $D$ fixed at $D=\bar{D} \pm \sigma_{D}$ and the other five parameters $p_{i}$ fixed at their nominal values. Notations are defined in the text.} 
\label{FigS_Uncert}
\end{center}
\end{figure}

The uncertainty and reproducibility of the fit procedure were first analyzed by fitting independent calibration linecuts while fixing the parameters $p_i$ to their nominal values $\bar{p_{i}}$. As an example, the histograms of the fit outcomes for $X=\{I_s,d\}$ are shown in Figure~\ref{FigS_Uncert}(a,b) for a set of 13 calibration linecuts recorded on the Ta/CoFeB/MgO sample with ND75c. From this statistic, we obtain $I_{s,\bar{p_{i}}}=926.3\pm2.8$~$\mu$A and $d_{\bar{p_{i}}}=122.9\pm0.7$~nm. Here the error bar is given by the standard deviation of the statistic. The relative uncertainty of the fit procedure is therefore given by $\epsilon_{d/{\rm fit}}=0.6 \%$ for the probe-sample distance and $\epsilon_{I_s/{\rm fit}}=0.3 \%$ for the product $I_s=M_st$.

We now estimate the relative uncertainty on the fit outcomes $(\epsilon_{d/p_{i}},\epsilon_{I_s/p_{i}})$ linked to each independent parameter $p_i$. For that purpose, the set of calibration linecuts was fitted with one parameter $p_i$ fixed at $p_i=\bar{p_{i}} \pm \sigma_{p_{i}}$, all the other five parameters remaining fixed at their nominal values. The resulting mean values of the fit parameters $X=\{d,I_s\}$ are denoted $X_{\bar{p_{i}} + \sigma_{p_{i}}}$ and $X_{\bar{p_{i}} - \sigma_{p_{i}}}$ and the relative uncertainty introduced by the errors on parameter $p_i$ is finally defined as 
\begin{equation} \label{partial_uncert}
\epsilon_{X/p_{i}}=\frac{X_{\bar{p_{i}} + \sigma_{p_{i}}}-X_{\bar{p_{i}} - \sigma_{p_{i}}}}{2X_{\bar{p_{i}} }} = \frac{\Delta_{X,p_i}}{2X_{\bar{p_{i}} }} \ .
\end{equation}
To illustrate the method, we plot in Figure~\ref{FigS_Uncert}(c,d) the histograms of the fit outcomes while changing the zero-field splitting parameter $D$ from $\bar{D}-\sigma_D$ to $\bar{D}+\sigma_D$. For this parameter, the relative uncertainties on $d$ and $I_s$ are $\epsilon_{d/D}=1.0\%$ and $\epsilon_{I_s/D}=1.6\%$. The same analysis was performed for all parameters $p_i$ and the corresponding uncertainties are summarized in Table~\ref{tab0}. The cumulative uncertainty is finally given by 
\begin{equation}
\epsilon_X=\sqrt{\epsilon^{2}_{X/{\rm fit}}+\sum_i \epsilon_{X/p_i}^2} \ ,
\label{EqUncert}
\end{equation}
where all errors are assumed to be independent. 

Following this procedure, we finally obtain $d=122.9 \pm 3.1$~nm and $M_s t= 926\pm26$~$\mu$A (or $M_s\approx0.926$ MA/m) for the Ta/CoFeB/MgO sample, and $d=119.0\pm 3.4$~nm and $M_s t= 671\pm18$~$\mu$A (or $M_s\approx1.12$ MA/m) for the Pt/Co/AlO$_x$ sample, in good agreement with the values reported elsewhere for similar samples~[\onlinecite{Vernier2014,Miron2010}].

\begin{table}[h!]
\begin{center}
{\large \bf (a)} {\bf Ta/CoFeB/MgO with ND75c}\\
\vspace{0.2cm}
\begin{tabular}{|c|c|c|c|c|}
\hline
parameter $p_i$ &  \ nominal value $\bar{p_i}$ \ & \ uncertainty $\sigma_{p_i}$ \ & $ \ \epsilon_{d/p_i}(\%) \ $ & $ \ \epsilon_{I_s/p_i}(\%) \ $  \\
\hline
 $w_c$ & 1500 nm & 30 nm & 1.8 & 2.0 \\
 $\delta d_m$ & $17$ nm & $2$ nm & 1.0 & 0.2 \\
 $\theta$ & $62^\circ$ & $2^\circ$ & 0.9 & 0.7 \\
 $\phi$ & $-25^\circ$ & $2^\circ$ & 0.2 & 1.2 \\
 $D$ & $2969.5$ MHz & $0.2$ MHz & 1.0 & 1.6 \\
 $E$ & $3.3$ MHz & $0.2$ MHz & 0.5 & 0.5 \\
 \hline
  \multicolumn{3}{|c|}{} &  &  \\
 \multicolumn{3}{|c|}{$\epsilon_X=\sqrt{\epsilon^{2}_{X/{\rm fit}}+\sum_i \epsilon_{X/p_i}^2}$} & 2.5 & 2.9 \\
  \multicolumn{3}{|c|}{} &  &  \\
   \hline
\end{tabular}
\end{center}

\begin{center}
{\large \bf (b)} {\bf Pt/Co/AlO$_x$ with ND79c}\\
\vspace{0.2cm}
\begin{tabular}{|c|c|c|c|c|}
\hline
parameter $p_i$ &  \ nominal value $\bar{p_i}$ \ & \ uncertainty $\sigma_{p_i}$ \ & $ \ \epsilon_{d/p_i}(\%) \ $ & $ \ \epsilon_{I_s/p_i}(\%) \ $  \\
\hline
 $w_c$ & 980 nm & 20 nm & 1.8 & 2.0 \\
 $\delta d_m$ & $25$ nm & $3$ nm & 1.6 & 0.4 \\
 $\theta$ & $87^\circ$ & $2^\circ$ & 0.2 & 0.1 \\
 $\phi$ & $23^\circ$ & $2^\circ$ & $<0.1$ & 1.4 \\
 $D$ & $2966.6$ MHz & $0.2$ MHz & 0.8 & 0.8 \\
 $E$ & $4.3$ MHz & $0.2$ MHz & $<0.1$ & $<0.1$ \\
 \hline
   \multicolumn{3}{|c|}{} &  &  \\
 \multicolumn{3}{|c|}{$\epsilon_X=\sqrt{\epsilon^{2}_{X/{\rm fit}}+\sum_i \epsilon_{X/p_i}^2}$} & 2.9 & 2.6 \\
  \multicolumn{3}{|c|}{} &  &  \\
   \hline
\end{tabular}
\caption{Summary of the uncertainty $\epsilon_{X/p_i}$ on the value of the fit parameter $X$ ($X=d$ and $X=I_s$) related to parameter $p_i$ for the experiments on Ta/CoFeB/MgO with ND75c (a) and on Pt/Co/AlO$_x$ with ND79c (b). The overall uncertainty $\epsilon_{X}$ is estimated with Eq.~(\ref{EqUncert}), assuming that all errors are independent. The standard deviations obtained from a series of 13 linecuts on Ta/CoFeB/MgO (resp. 9 linecuts on Pt/Co/AlO$_x$) are $\epsilon_{d/{\rm fit}}=0.6 \%$ and $\epsilon_{I_s/{\rm fit}}=0.3 \%$ (resp. $\epsilon_{d/{\rm fit}}=1.4 \%$ and $\epsilon_{I_s/{\rm fit}}=0.5 \%$).}
\label{tab0}
\end{center}
\end{table}

\subsection{Micromagnetic calculations} \label{simulations}

While the calibration linecuts were fitted with analytic formulas, the predictions of the stray field above the DWs were obtained using micromagnetic calculations in order to accurately describe the DW fine structure. We first used the micromagnetic OOMMF software [\onlinecite{oommf,Rohart2013}] to obtain the equilibrium magnetization of the structure. For the Ta/CoFeB/MgO sample, the nominal values used in OOMMF are: anisotropy constant $K=5.9\cdot 10^5$ J/m$^3$ (obtained from the measured effective anisotropy field of 107 mT [\onlinecite{Devolder2013}]), exchange constant $A=20$ pJ/m, film thickness $t=1$ nm, stripe width $w=1500$ nm, cell size $2.5\times2.5\times1$ nm$^3$. For the Pt/Co/AlO$_x$ sample, we used: $K=1.3\cdot 10^6$ J/m$^3$ (measured effective anisotropy field of 920 mT), $A=18$ pJ/m, $t=0.6$ nm, $w=470$ nm, cell size $2.5\times2.5\times0.6$ nm$^3$. The saturation magnetization $M_s$ was obtained from the product $M_st$ determined from calibration linecuts [cf. Section \ref{calib}]. 

\begin{figure}[t]
\begin{center}
\includegraphics[width=1\textwidth]{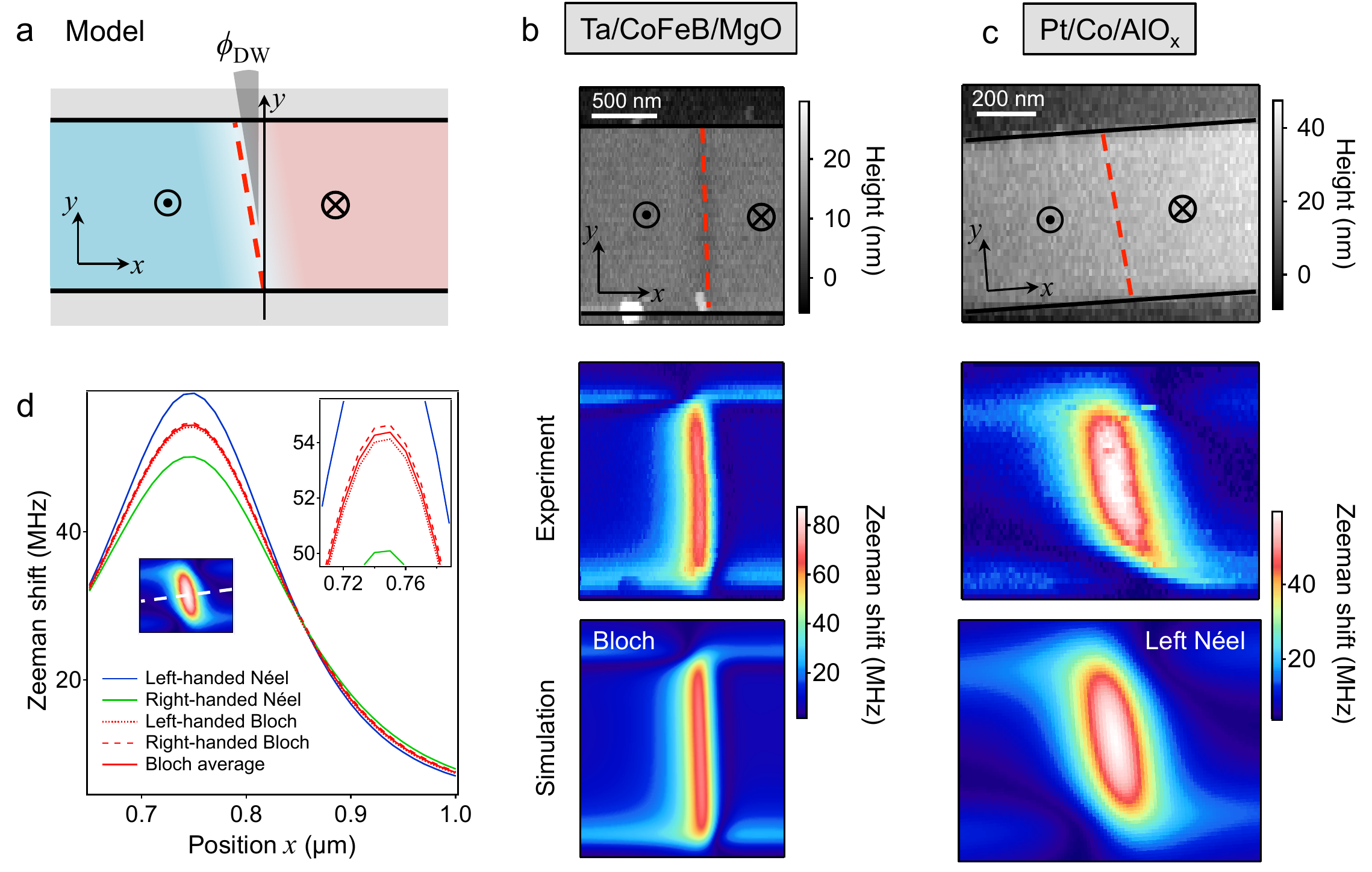}
\caption{(a) The DW is assumed to be straight with a tilt angle $\phi_{\rm DW}$ with respect to the $y$ axis, perpendicular to the wire's long axis. (b,c) AFM image (top panel), Zeeman shift image (middle panel) and associated simulation (bottom panel) corresponding to the DW studied in (b) Fig. 2 and (c) Fig. 4 of the main paper. The simulation assumes a straight DW with $\phi_{\rm DW}=2^\circ$ and $\psi=\pi/2$ in (b), and $\phi_{\rm DW}=6^\circ$ and $\psi=\pi$ in (c). (d) Linecuts taken from the simulation of (c), illustrating the small effect of the chirality of the Bloch DW. Near the maximum, the field is changed by $\pm0.5\%$ with respect to the mean value. In the case of (b), the change is even smaller ($\pm0.3\%$).}
\label{FigS_DWtilt}
\end{center}
\end{figure}

We considered a straight DW with a tilt angle $\phi_{\rm DW}$ with respect to the $y$ axis [Fig.~\ref{FigS_DWtilt}(a)]. As illustrated in Figs.~\ref{FigS_DWtilt}(b) and \ref{FigS_DWtilt}(c), this angle was directly inferred from the Zeeman shift images, leading to $\phi_{\rm DW}\approx2\pm1^\circ$ for the DW studied in Fig. 2 of the main paper, and $\phi_{\rm DW}\approx6\pm2^\circ$  for the DW studied in Fig. 4 of the main paper. The uncertainty on $\phi_{\rm DW}$ enables us to account for the fact that the DW is not necessarily rigorously straight. This point will be discussed in Section \ref{uncert}. 

The calculation of the stray field was then performed with four different initializations of the DW magnetization: (i) right-handed Bloch, (ii) left-handed Bloch, (iii) right-handed N\'eel and (iv) left-handed N\'eel. To stabilize the N\'eel configuration, the DMI at one of the interfaces of the ferromagnet was added, as described in Ref.~[\onlinecite{Rohart2013}]. The value of the DMI parameter was set to $|D_{\rm DMI}|=0.5$ mJ/m$^2$, which is large enough to fully stabilize a N\'eel DW. The additional consequences of a stronger DMI will be discussed in Section \ref{DMItilt}.

Once the equilibrium magnetization was obtained, the stray field distribution ${\bf B}(x,y)$ was calculated at the distance $d$ by summing the contribution of all cells. Knowing the projection axis ($\theta$,$\phi$), we finally calculate the Zeeman shift map $\Delta f_{\rm NV}(x,y)$ by diagonalizing the NV center's Hamiltonian [cf. Section \ref{NVcharac}]. Under the conditions of Figs. 2 and 4 of the main paper, the difference of stray field near the maximum between left-handed and right-handed Bloch DWs is predicted to be $< 0.5\%$ [Fig.~\ref{FigS_DWtilt}(d)]. Since this is much smaller than the standard error [cf. Section \ref{uncert}], we plotted the mean of these two cases, which is simply referred to as a Bloch DW, and added the deviation induced by the two possible chiralities to the displayed standard error.

\subsection{Uncertainties on the DW stray field predictions} \label{uncert}

In this Section, we analyze how the uncertainties on the preliminary measurements affect the final predictions of the Zeeman shift above the DW. To keep the analysis simple and insightful, we use the approximate analytic expressions of the stray field of an infinitely long DW [Eqs. (1), (2) and (3) of the main paper]. Furthermore, we focus our attention on the positions where the DW stray field is maximum, since this is what provides information about the DW nature [see Figs. 1(c) and 1(d) of the main paper]. Finally, we use the approximation $\Delta f_{\rm NV}\approx g\mu_B B_{\rm NV,\parallel}/h$ [cf. Section \ref{NVcharac}], which is quite accurate near the stray field maximum and allows us to consider the magnetic field $B_{\rm NV,\parallel}$ rather than the Zeeman shift $\Delta f_{\rm NV}$. For clarity the subscript $\parallel$ will be dropped and the projected field will be simply denoted $B_{\rm NV}$. 

\begin{figure}[h!]
\begin{center}
\includegraphics[width=0.85\textwidth]{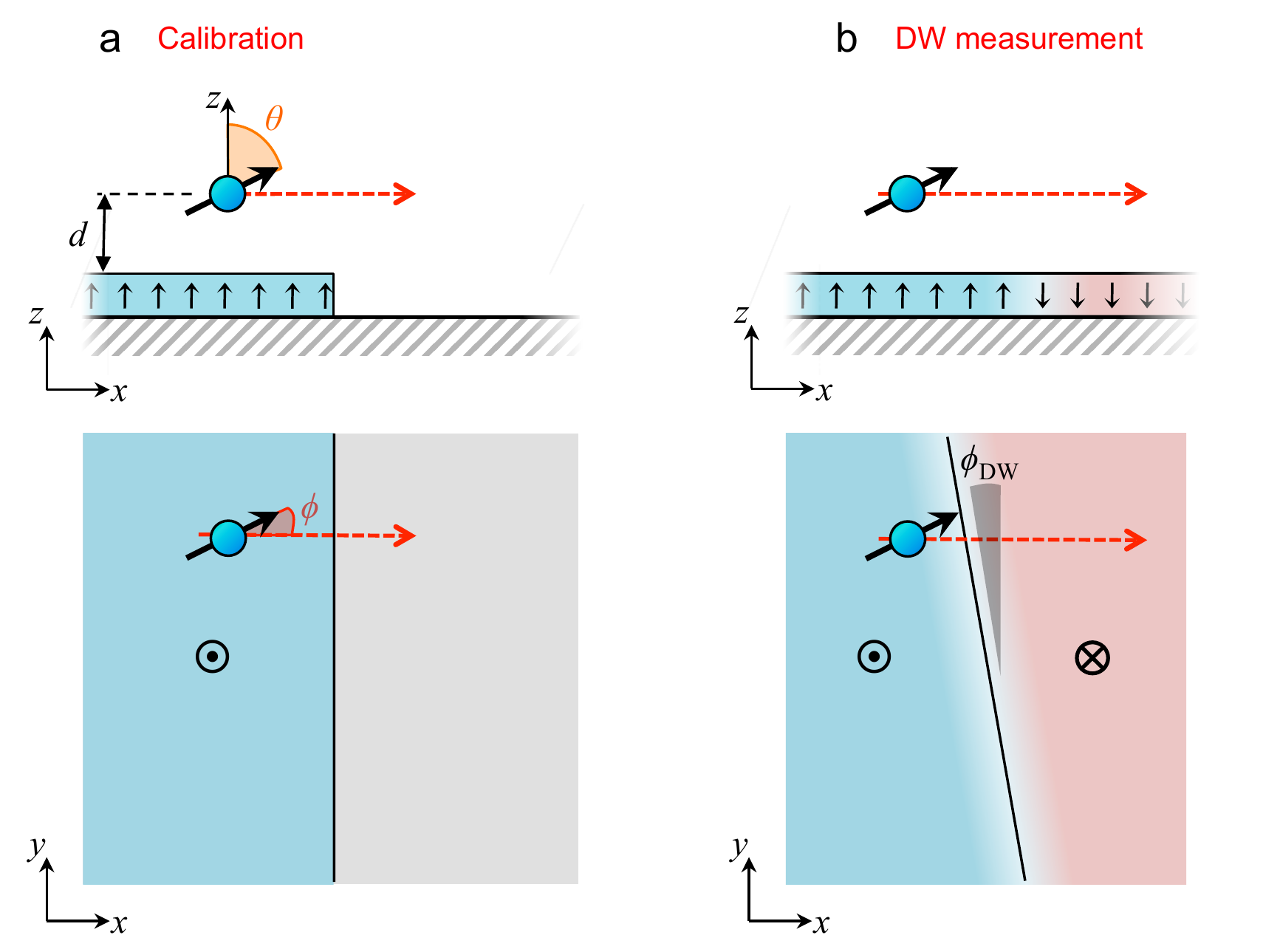}
\caption{To estimate the uncertainty in the DW stray field prediction, we analyze how the error on a calibration measurement above an edge (a) and on other parameters translates into an error on the DW field (b). The calibration edge defines the ($xyz$) axis system. The DW is assumed to be infinitely long, with its plane tilted by an angle $\phi_{\rm DW}$ with respect to the $(yz)$ plane. The angle $\psi$ defines the rotation of the in-plane magnetization of the DW with respect to the DW normal. Top panels: side view; Bottom panels: top view.} 
\label{FigS_uncert}
\end{center}
\end{figure}

\subsubsection{Out-of-plane contribution ${\bf B}^\perp$}

Let us first consider the out-of-plane contribution to the DW stray field, ${\bf B}^\perp(x)$. The stray field components above the DW can be written, in the $(xyz)$ axis system (Fig. \ref{FigS_uncert}), as
\begin{equation}
\begin{dcases} 
B_x^\perp(x)=\frac{\mu_0M_st}{\pi}\frac{d \cos\phi_{\rm DW} }{\left[ (x-x_{\rm DW})\cos\phi_{\rm DW}\right]^2+d^2} \\
B_y^\perp(x)=\frac{\mu_0M_st}{\pi}\frac{d \sin\phi_{\rm DW} }{\left[ (x-x_{\rm DW})\cos\phi_{\rm DW}\right]^2+d^2} \\
B_z^\perp(x)=-\frac{\mu_0M_st}{\pi}\frac{ (x-x_{\rm DW})\cos\phi_{\rm DW}}{\left[ (x-x_{\rm DW})\cos\phi_{\rm DW}\right]^2+d^2}~,
\end{dcases}
\label{eq:B_Bloch} 
\end{equation}   
where $x_{\rm DW}$ is the position of the DW (for a given $y$). This is simply twice the stray field above an edge [see Eq. (\ref{eq:Bedge})] expressed in a rotated coordinate system. The projection along the NV center's axis is 
\begin{eqnarray} 
B_{\rm NV}^\perp(x) & = & \left \vert \sin\theta\cos\phi B_x^\perp(x)+\sin\theta\sin\phi B_y^\perp(x)+\cos\theta B_z^\perp(x) \right\vert
\\
& = & \frac{\mu_0M_st}{\pi}\frac{1}{\left[ (x-x_{\rm DW})\cos\phi_{\rm DW}\right]^2+d^2} \left\vert d\sin\theta\cos(\phi-\phi_{\rm DW})-(x-x_{\rm DW})\cos\phi_{\rm DW}\cos\theta \right\vert~.
\label{eq:BNV_DW}
\end{eqnarray}
We now link $B_{\rm NV}^\perp(x)$ to the calibration measurement. For simplicity, we consider only one of the two edges of the calibration stripe, {\it e.g.} the edge at $x=0$. 
We can thus write the projected field above the edge, at a distance $d$, as 
\begin{eqnarray} 
B_{\rm NV}^{\rm edge}(x) & = &  \left \vert \sin\theta\cos\phi B_x^{\rm edge}(x)+\sin\theta\sin\phi B_y^{\rm edge}(x)+\cos\theta B_z^{\rm edge}(x) \right\vert \\
& = & \frac{\mu_0 M_st}{2\pi}\frac{1}{x^2+d^2} \left\vert d\sin\theta\cos\phi-x\cos\theta \right\vert~.
\label{eq:BNV_edge}
\end{eqnarray}
Comparing Eqs. (\ref{eq:BNV_DW}) and (\ref{eq:BNV_edge}), one finds the relation
\begin{eqnarray}  \label{eq:BNV_relation}
B_{\rm NV}^\perp\left(\frac{x}{\cos\phi_{\rm DW}}+x_{\rm DW}\right) & = & 2 B_{\rm NV}^{\rm edge}(x)\Theta_{d,\theta,\phi,\phi_{\rm DW}}(x)~,
\end{eqnarray}
where we define
\begin{equation} \label{ThetaG}
\Theta_{d,\theta,\phi,\phi_{\rm DW}}(x) = \left\vert\frac{d\sin\theta\cos(\phi-\phi_{\rm DW})-x\cos\theta}{d\sin\theta\cos\phi-x\cos\theta} \right\vert~.
\end{equation} 
 
Since $B_{\rm NV}^{\rm edge}(x)$ is experimentally measured, in principle one can use Eq. (\ref{eq:BNV_relation}) to predict $B_{\rm NV}^{\perp}(x)$ by simply evaluating the function $\Theta_{d,\theta,\phi,\phi_{\rm DW}}(x)$ as defined by Eq. (\ref{ThetaG}). As $\phi_{\rm DW}\sim 0$ implies $\Theta_{d,\theta,\phi,\phi_{\rm DW}}(x)\sim 1$, it comes that, in a first approximation, $B_{\rm NV}^{\perp}(x)$ can be obtained without the need for precise knowledge of any parameter. In other words, the calibration measurement, performed under the same conditions as for the DW measurement, allows us to accurately predict the DW field even though those conditions are not precisely known. This is the key point of our analysis. 

Strictly speaking, $\Theta_{d,\theta,\phi,\phi_{\rm DW}}(x)$, hence $B_{\rm NV}^{\perp}(x)$, does depend on some parameters as soon as $\phi_{\rm DW}\neq 0$, namely on $\{q_i\}=\{d,\theta,\phi,\phi_{\rm DW}\}$. To get an insight into how important the knowledge of $\{q_i\}$ is, we need to examine how sensitive $\Theta_{d,\theta,\phi,\phi_{\rm DW}}(x)$ is with respect to errors on $\{q_i\}$. Owing to the sine and cosine functions in Eq. (\ref{ThetaG}), the smallest sensitivity to parameter variations (vanishing partial derivatives) is achieved when either (i) $\theta\sim 0$ (projection axis perpendicular to the sample plane) or (ii) $\theta\sim \pi/2$ (projection axis parallel to the sample plane) combined with $\phi\sim 0$ and $\phi-\phi_{\rm DW}\sim 0$. However, case (i) cannot be achieved in our experiment, because the out-of-plane RF field cannot efficiently drive ESR of a spin pointing out-of-plane. We therefore target case (ii), that is, $\theta\sim \pi/2$ and $\phi-\phi_{\rm DW}\sim 0$. For that purpose, we use a calibration edge that is as parallel to the DW as possible ($\phi_{\rm DW}\rightarrow0$) and we seek to have a projection axis that is as perpendicular to the DW plane as possible ($\theta\rightarrow\pi/2$ and $\phi\rightarrow0$). This is why we employ two perpendicular wires for the calibration and the DW measurements, respectively [cf. Section \ref{samples}]. Conversely, in the worst case of $\phi_{\rm DW}\sim\pi/2$ (calibration edge perpendicular to the DW) with $\theta\sim\pi/2$, one would have $\Theta_{d,\theta,\phi,\phi_{\rm DW}}(x)\sim \phi-\phi_{\rm DW}$, directly proportional to the errors on $\phi$ and $\phi_{\rm DW}$.  

To be more quantitative, we use Eq. (\ref{eq:BNV_relation}) to express the uncertainty on the prediction $B_{\rm NV}^{\perp}(x)$ as a function of the uncertainties on the various quantities, which gives
\begin{equation}
\epsilon_{B^\perp}=\sqrt{\epsilon^2_{B^{\rm edge}}+\sum_i \epsilon_{\Theta/q_i}^2} \ .
\label{EqUncert2}
\end{equation}
Here, $\epsilon_{B^{\rm edge}}$ is given by the measurement error of $B_{\rm NV}^{\rm edge}(x)$, whereas $\epsilon_{\Theta/q_i}$ is the uncertainty on $\Theta_{\lbrace q_i \rbrace}$ introduced by the error on the parameter $q_i\in\lbrace d,\theta,\phi,\phi_{\rm DW}\rbrace$, the other parameters being fixed at their nominal values, as defined by 
\begin{equation} \label{partial_uncert2}
\epsilon_{\Theta/q_{i}}=\frac{\Theta_{\bar{q_{i}} + \sigma_{q_{i}}}-\Theta_{\bar{q_{i}} - \sigma_{q_{i}}}}{2\Theta_{\bar{q_{i}} }} \ .
\end{equation}
The results are summarized in Table \ref{tab2} for the cases considered in Figs. 2 (Ta/CoFeB/MgO sample) and 4 (Pt/Co/AlO$_x$ sample) of the main paper. $\epsilon_{\Theta/q_{i}}$ is evaluated for $x=x_{\rm max}$, which is the position where the field $B_{\rm NV}^{\perp}(x)$ is maximum. It can be seen that the dominating source of uncertainty, though small ($\approx 1\%$), is the error on $\phi_{\rm DW}$, while the errors on $d$, $\theta$ and $\phi$ have a negligible impact.

In practice, to obtain the theoretical predictions shown in the main paper and in Fig. \ref{FigS_DWtilt}, we do not use explicitly Eq. (\ref{eq:BNV_relation}), but rather use the set of parameters $\left\lbrace I_s,d,\theta,\phi \right\rbrace$ determined following the calibration step, and put it into the stray field computation [cf. Section \ref{simulations}]. This allows us to simulate more complex structures than the idealized infinitely long DW considered above [Fig. \ref{FigS_uncert}(b)], in particular the finite-width wires studied in this work. However, we stress that, as far as the uncertainties are concerned, this is completely equivalent to using Eq. (\ref{eq:BNV_relation}), since $B_{\rm NV}^{\rm edge}(x)$ is fully characterized by the set $\left\lbrace I_s,d,\theta,\phi \right\rbrace$ [cf. Section \ref{calib}]. The main difference comes from the influence of the edges of the wire, of width $w$, on the DW stray field. The standard error $\sigma_w$ then translates into a relative error $\epsilon_{B^\perp/w}$ on the DW field $B_{\rm NV}^{\perp}$. For the Ta/CoFeB/MgO sample, $w=1500\pm30$ nm, which gives a negligible error $\epsilon_{B^\perp/w}<0.1\%$ for the field calculated at the center of the stripe. For the Pt/Co/AlO$_x$ sample, the stripe is narrower, $w=470\pm20$ nm, leading to $\epsilon_{B^\perp/w}=0.9\%$. The overall uncertainty on the prediction $B_{\rm NV}^{\perp}$, for a DW confined in a wire, then becomes
\begin{equation}
\epsilon_{B^\perp}=\sqrt{\epsilon_{B^\perp/w}^2+\epsilon^{2}_{B^{\rm edge}}+\sum_i \epsilon_{\Theta/q_i}^2} \ .
\label{EqUncert3}
\end{equation}
The overall errors are indicated in Table \ref{tab2}. For Ta/CoFeB/MgO (Fig. 2 of the main paper ), the overall standard error is found to be $\approx 1.5\%$, whereas for Pt/Co/AlO$_x$ (Fig. 4) it is $\approx 2.1\%$, in both cases much smaller than the difference between Bloch and N\'eel DW configurations.

\begin{table}[h!]
\begin{center}
{\large \bf (a)} {\bf Ta/CoFeB/MgO with ND75c}\\
\vspace{0.5cm}
\begin{tabular}{|c|c|c|c|}
\hline
parameter $q_i$ &  \ nominal value $\bar{q_i}$ \ & \ uncertainty $\sigma_{q_i}$ \ & $ \ \epsilon_{\Theta/q_i}(\%) \ $  \\
\hline
 $d$ & $123$ nm & $3$ nm & $<0.1$ \\
 $\theta$ & $62^\circ$ & $2^\circ$ & $<0.1$ \\
 $\phi$ & $-25^\circ$ & $2^\circ$ & 0.2 \\
 $\phi_{\rm DW}$ & $2^\circ$ & $1^\circ$ & 1.1 \\
 \hline
  \multicolumn{3}{|c|}{} &    \\
 \multicolumn{3}{|c|}{$\epsilon_{B^\perp}=\sqrt{\epsilon_{B^\perp/w}^2+\epsilon^{2}_{B^{\rm edge}}+\sum_i \epsilon_{\Theta/q_i}^2}$} & 1.5  \\
  \multicolumn{3}{|c|}{} &    \\
   \hline
\end{tabular}
\end{center}

\begin{center}
{\large \bf (b)} {\bf Pt/Co/AlO$_x$ with ND79c}\\
\vspace{0.5cm}
\begin{tabular}{|c|c|c|c|}
\hline
parameter $q_i$ &  \ nominal value $\bar{q_i}$ \ & \ uncertainty $\sigma_{q_i}$ \ & $ \ \epsilon_{\Theta/q_i}(\%) \ $  \\
\hline
 $d$ & $118$ nm & $4$ nm & $<0.1$ \\
 $\theta$ & $87^\circ$ & $2^\circ$ & $<0.1$ \\
 $\phi$ & $23^\circ$ & $2^\circ$ & 0.4 \\
 $\phi_{\rm DW}$ & $6^\circ$ & $2^\circ$ & 1.1 \\
 \hline
   \multicolumn{3}{|c|}{} &    \\
 \multicolumn{3}{|c|}{$\epsilon_{B^\perp}=\sqrt{\epsilon_{B^\perp/w}^2+\epsilon^{2}_{B^{\rm edge}}+\sum_i \epsilon_{\Theta/q_i}^2}$} & 2.1  \\
  \multicolumn{3}{|c|}{} &    \\
   \hline
\end{tabular}
\caption{Summary of the uncertainty $\epsilon_{\Theta/q_i}$ on the value of $\Theta$ related to parameter $q_i$ for the experiments on Ta/CoFeB/MgO with ND75c (a) and on Pt/Co/AlO$_x$ with ND79c (b). The overall uncertainty $\epsilon_{B^\perp}$ is estimated with Eq.~(\ref{EqUncert3}), assuming that all errors are independent. The relative error on the calibration field $B_{\rm NV}^{\rm edge}(x)$ is estimated to be $\epsilon_{B^{\rm edge}}\approx 1.0\%$ in (a) and $\epsilon_{B^{\rm edge}}\approx 1.5\%$ in (b). The effect of the stripe width uncertainty leads to an additional error $\epsilon_{B^\perp/w}<0.1\%$ in (a) and $\epsilon_{B^\perp/w}=0.9\%$ in (b).}
\label{tab2}
\end{center}
\end{table}

\subsubsection{In-plane contribution ${\bf B}^\parallel$}

According to Eq. (2) of the main paper, the in-plane contribution to the DW stray field, ${\bf B}^\parallel(x)$, is proportional to $I_s$ and to the DW width $\Delta_{\rm DW}=\sqrt{A/K_{\rm eff}}$, where $A$ is the exchange constant and $K_{\rm eff}$ the effective anisotropy constant. The values of $A$ reported in the literature for Co and CoFeB thin films range from 10 to 30 pJ/m (see {\it e.g.} Refs. [\onlinecite{Yamanouchi2011,Eyrich2012}]). Based on this range, we deduced a range for $\Delta_{\rm DW}$, namely 15-25 nm for the Ta/CoFeB/MgO sample and 4.4-7.6 nm for the Pt/Co/AlO$_x$ sample. This amounts to a relative variation $\frac{\sigma_{\Delta_{\rm DW}}}{\Delta_{\rm DW}}\approx25\%$ around the mid-value of $\Delta_{\rm DW}$. Thus, $\epsilon_{B^\parallel}$ is dominated by the uncertainty on the DW width, that is, $\epsilon_{B^\parallel}\approx\frac{\sigma_{\Delta_{\rm DW}}}{\Delta_{\rm DW}}\approx25\%$. All other errors can be neglected in comparison. In the simulations [cf. Section \ref{simulations}], we used the value of $A$ that gives the mid-value of $\Delta_{\rm DW}$, that is $A=20$ pJ/m for Ta/CoFeB/MgO ($\Delta_{\rm DW}=20$ nm) and $A=18$ pJ/m for Pt/Co/AlO$_x$ ($\Delta_{\rm DW}=6.0$ nm). 

For an arbitrary angle $\psi$ of the in-plane magnetization of the DW, the projected stray field writes
\begin{eqnarray}
B_{\rm NV}^{\psi}(x) & = & B_{\rm NV}^{\perp}(x) + \cos\psi B_{\rm NV}^{\parallel}(x)~,
\end{eqnarray}
where it is assumed that $|B_{\rm NV}^{\parallel}|<|B_{\rm NV}^{\perp}|$. We deduce the expression of the absolute uncertainty for $B_{\rm NV}^{\psi}$
\begin{eqnarray}
\sigma_{B^{\psi}} =\sqrt{ \sigma_{B^\perp}^2 + \cos^2 \psi \sigma_{B^\parallel}^2}~,
\end{eqnarray}
where $\sigma_{B^\perp}=\epsilon_{B^\perp}B_{\rm NV}^{\perp}$ and $\sigma_{B^\parallel}=\epsilon_{B^\parallel}B_{\rm NV}^{\parallel}$. This is how the confidence intervals shown in Figs. 2 and 4 of the main paper were obtained. Finally, the confidence intervals for $\cos\psi$ were defined as the values of $\cos\psi$ such that the data points remain in the interval $[B_{\rm NV}^{\psi}-\sigma_{B^{\psi}}~;~B_{\rm NV}^{\psi}+\sigma_{B^{\psi}}]$. The interval for the DMI parameter was deduced using the relation~[\onlinecite{Thiaville2012}]
\begin{eqnarray}
D_{\rm DMI} =\frac{2\mu_0M_s^2t\ln 2}{\pi^2} \cos\psi~,
\end{eqnarray}
which holds for an up-down DW provided that $|\cos\psi|<1$.

\subsection{Effects of a large DMI constant} \label{DMItilt}

So far, we have only considered, for simplicity and to avoid introducing additional parameters, the effect of DMI on the angle $\psi$ of the in-plane DW magnetization. In doing so, two other effects of DMI have been neglected: (i) the DMI induces a rotation of the magnetization near the edges of the ferromagnetic structure~[\onlinecite{Rohart2013}] and (ii) the DW profile in the presence of DMI slightly deviates from the profile $M_z(x)=-M_s \tanh(x/\Delta_{\rm DW})$~[\onlinecite{Thiaville2012}]. The first (second) effect modifies the stray field above the calibration stripe (above the DW). Here we quantify these effects for the case of Pt/Co/AlO$_x$, for which the DMI is expected to be strong. 

\begin{figure}[htb!]
\begin{center}
\includegraphics[width=0.99\textwidth]{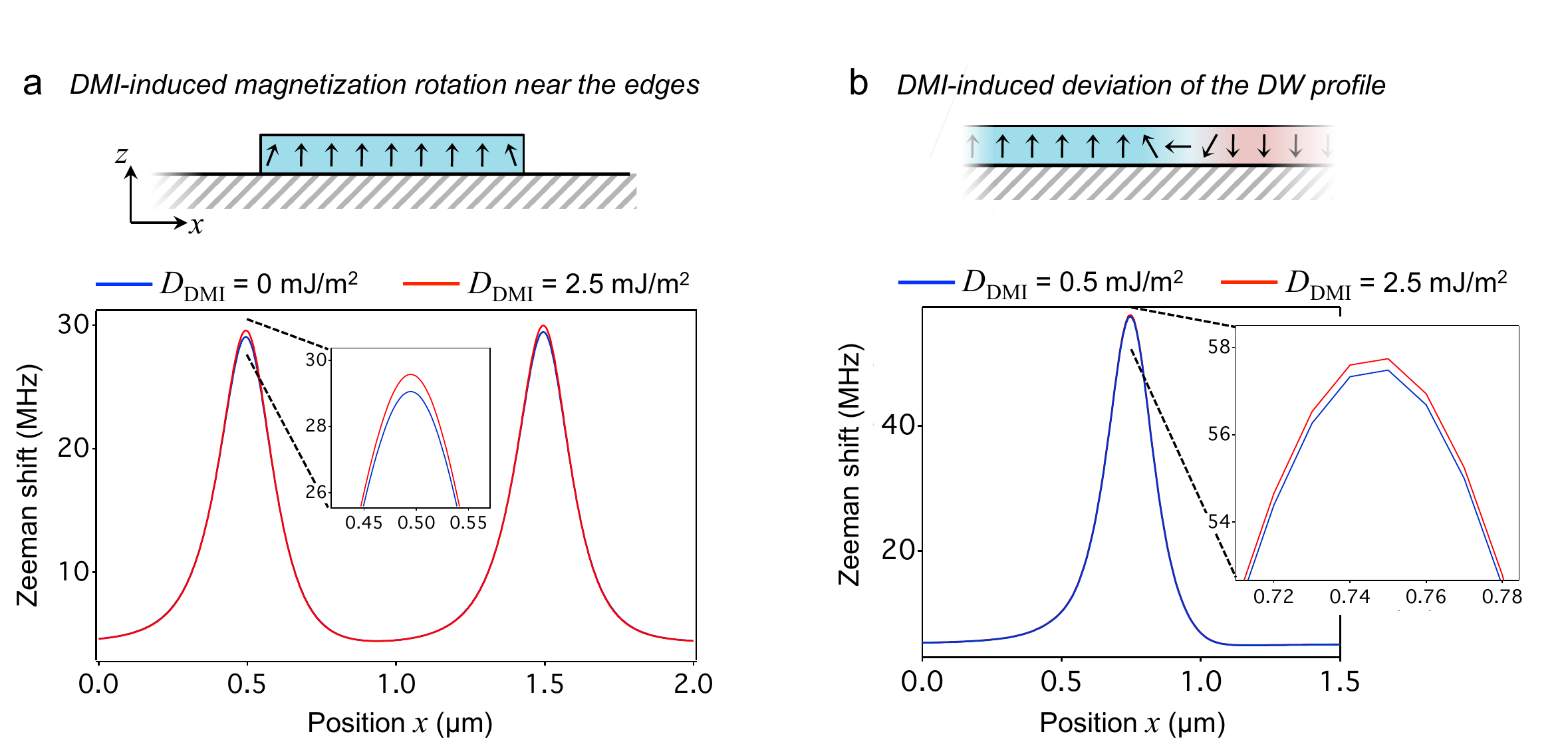}
\caption{(a) In the presence of a strong DMI, the magnetization deviates from the out-of-plane direction near the edges of the stripe. The plot shows the Zeeman shift calculated under similar conditions as in Fig. \ref{FigS_Linecuts}(b), for two different values of  $D_{\rm DMI}$. (b) The DMI also makes the DW profile deviate from the profile $M_z(x)=-M_s \tanh(x/\Delta_{\rm DW})$. The plot shows the Zeeman shift calculated under similar conditions as in Fig. 4 of the main paper, for two different values of  $D_{\rm DMI}$.}
\label{FigS_DMIeffects}
\end{center}
\end{figure}

Recently, Martinez {\it et al.} have estimated that a value $D_{\rm DMI}=-2.4$ mJ/m$^2$ associated with the spin Hall effect would quantitatively reproduce current-induced DW velocity measurements in Pt/Co/AlO$_x$~[\onlinecite{Martinez2013}]. On the other hand, Pizzini {\it et al.} have inferred a similar value of $D_{\rm DMI}=-2.2$ mJ/m$^2$ from field-dependent DW nucleation experiments~[\onlinecite{Pizzini2014}]. This is $\approx70\%$ of the threshold value $D_c$ above which the DW energy becomes negative and a spin spiral develops. Taking $D_{\rm DMI}=-2.5$ mJ/m$^2$, we predict that the magnetization rotation at the edges reaches $\approx 20^\circ$~[\onlinecite{Rohart2013}]. As a result, the field maximum above the edge is increased by $\approx 1.8\%$, under the conditions of Fig. \ref{FigS_Linecuts}(b) [Fig. \ref{FigS_DMIeffects}(a)].  This is of the order of our measurement error, so that this DMI-induced magnetization rotation cannot be directly detected in our experiment. In fitting the data of Fig. \ref{FigS_Linecuts}(b), the outcome for $I_s$ and $d$ is changed by a similar amount: we found $d=119.0\pm 3.4$~nm and $I_s= 671\pm18$~$\mu$A without DMI, as compared with $d=121.0\pm 3.4$~nm and $I_s= 670\pm17$~$\mu$A if $D_{\rm DMI}=-2.5$ mJ/m$^2$ is included. The difference is below the uncertainty, therefore it does not affect the interpretation of the data measured above the DW. 

To quantify the second effect, we performed the OOMMF calculation with two different values of $D_{\rm DMI}$ that stabilize a left-handed N{\'e}el DW: $D_{\rm DMI}=-0.5$ mJ/m$^2$, as used for the simulations shown in the main paper, and $D_{\rm DMI}=-2.5$ mJ/m$^2$. The stray field calculations, under the same conditions as in Fig. 4 of the main paper, show an increase of the field maximum by $\approx0.5\%$ for the stronger DMI [Fig. \ref{FigS_DMIeffects}(b)]. Again, this is well below the uncertainty [cf. Section \ref{uncert}].

Besides, it is worth pointing out that these two effects tend to compensate each other, since the first one tends to increase the estimated distance $d$, thereby decreasing the predicted DW field, while the second one tends instead to increase the predicted DW field. Overall, we conclude that neglecting the additional effects of DMI provides predictions for the N{\'e}el DW stray field that are correct within the uncertainty, even with a DMI constant as large as $70\%$ of $D_c$. We note finally that the predictions for the Bloch case, as plotted in Fig. 2 and 4 of the main paper, are anyway not affected by the above considerations, since the Bloch case implies no DMI at all.

\end{document}